\pdfoutput=1

\documentclass[twocolumn]{emulateapj}
\usepackage{natbib}
\usepackage{amsmath}
\usepackage{verbatim}

\begin{document}

\title{\textbf{Radar Imaging and Physical Characterization of Near-Earth Asteroid
(162421) 2000 ET70}} \date{} 
\author{S. P. Naidu\altaffilmark{1}, J.~L. Margot\altaffilmark{1,2}, M. W. Busch\altaffilmark{1,6},
  P. A. Taylor\altaffilmark{3}, M. C. Nolan\altaffilmark{3}, 
  M. Brozovic\altaffilmark{4}, L.~A.~M. Benner\altaffilmark{4}, J.~D.~Giorgini\altaffilmark{4}, 
  C. Magri\altaffilmark{5}}

\altaffiltext{1}{Department of Earth and Space Sciences, University of California, Los Angeles, CA 90095, USA}
\altaffiltext{2}{Department of Physics and Astronomy, University of California, Los Angeles, CA 90095, USA}
\altaffiltext{3}{Arecibo Observatory, HC3 Box 53995, Arecibo, PR 00612, USA}
\altaffiltext{4}{Jet Propulsion Laboratory, California Institute of Technology, Pasadena, CA 91109-8099, USA}
\altaffiltext{5}{University of Maine at Farmington, 173 High Street, Preble Hall, Farmington, ME 04938, USA}
\altaffiltext{6}{Now at National Radio Astronomy Observatory, 1003 Lopezville Road, Socorro, NM 87801, USA}

\begin{abstract}
We observed near-Earth asteroid (162421) 2000~ET70 using the Arecibo
and Goldstone radar systems over a period of 12 days during its close
approach to the Earth in February 2012. We obtained continuous wave
spectra and range-Doppler images with range resolutions as fine as
15~m. Inversion of the radar images yields a detailed shape model with
an effective spatial resolution of 100~m.  The asteroid has overall
dimensions of 2.6~km $\times$ 2.2~km $\times$ 2.1~km (5\%
uncertainties) and a surface rich with kilometer-scale ridges and
concavities.  This size, combined with absolute magnitude
measurements, implies an extremely low albedo ($\sim$2\%).  It is a
principal axis rotator and spins in a retrograde manner with a
sidereal spin period of 8.96~$\pm$~0.01 hours.  In terms of
gravitational slopes evaluated at scales of 100 m, the surface seems
mostly relaxed with over 99\% of the surface having slopes less than
30$^\circ$, but there are some outcrops at the north pole that may
have steeper slopes.  Our precise measurements of the range and
velocity of the asteroid, combined with optical astrometry, enables
reliable trajectory predictions for this potentially hazardous
asteroid in the interval 460-2813.
\end{abstract}


\section{Introduction}

Radar astronomy is arguably the most powerful Earth-based technique
for characterizing the physical properties of near-Earth asteroids
(NEAs). Radar observations routinely provide images with decameter
spatial resolution.  These images can be used to obtain accurate
astrometry, model shapes, measure near-surface radar scattering
properties, and investigate many other physical properties
(e.g. sizes, spin states, masses, densities).  Radar observations have
led to the discovery of asteroids exhibiting non-principal axis
rotation~\citep[e.g.,][]{ostro95,benn02}, binary and triple
NEAs~\citep[e.g.,][]{margot02,ostro06,shepard06,nola08dps,broz11}, and
contact binary asteroids~\citep[e.g.,][]{huds94,benn06,broz10}.
Radar-derived shapes and spins have been used to investigate various
physical processes (Yarkovsky, YORP, BYORP, tides, librations,
precession, etc.)  that are important to the evolution of
NEAs~\citep[e.g.,][]{ches03,nuge12yark,lowr07,tayl07,ostro06,scheeres06,tayl11,fang11triples,fang12spinorbit}.

Here we present the radar observations and detailed physical
characterization of NEA (162421) 2000~ET70. This Aten asteroid
(a=0.947 AU, e=0.124, i=22.3$^{\circ}$) was discovered on March 8,
2000 by the Lincoln Near-Earth Asteroid Research (LINEAR) program in
Socorro, New Mexico.  Its absolute magnitude was reported to be
18.2~\citep{whiteley01} which, for typical optical albedos between 0.4
and 0.04, suggests a diameter between 0.5 and 1.5 km.  Recent analysis
of the astrophotometry yields absolute magnitude values that are
comparable~\citep{will12}.  \citet{alvarez12} obtained a lightcurve of
the asteroid during its close approach to the Earth in February
2012. They reported a lightcurve period of 8.947~$\pm$~0.001 hours and
a lightcurve amplitude of 0.60~$\pm$~0.07.  Using visible photometry,
\citet{whiteley01} classified 2000~ET70 as an X-type asteroid in
the~\citet{tholen84} taxonomy.  The Tholen X class is a degenerate
group of asteroids consisting of E, M, and P classes, which are
distinguished by albedo. Mike Hicks (personal communication) used
visible spectroscopy and indicated that his observations best matched
a C-type or possibly an E-type.  Using spectral observations covering
a wavelength range of 0.8 to 2.5~${\rm\mu m}$ in addition to the
visible data, Ellen Howell (personal communication) classified it as
Xk in the taxonomic system of Bus-DeMeo~\citep{demeo09}.

\section{Observations and data processing}
\label{sec:obs}
We observed 2000~ET70 from February 12, 2012 to February 17, 2012
using the Arecibo S-band (2380 MHz, 13 cm) radar and from February 15,
2012 to February 23, 2012 using the Goldstone X-band (8560 MHz, 3.5
cm) radar. The asteroid moved $\sim$74~degrees across the sky during
this time and it came closest to Earth on February 19 at a distance of
$\sim$0.045~Astronomical Units (AU).  We
observed it again in August 2012 when it made another close approach
to Earth at a distance of $\sim$0.15 AU.

Radar observing involved transmitting a radio wave for approximately
the round-trip light-time (RTT) to the asteroid, $\sim$46 seconds at
closest approach, and then receiving the echo reflected back from the
target for a comparable duration. Each transmit-receive cycle is
called a {\em run}. On each day we carried out runs with a
monochromatic continuous wave (CW) to obtain Doppler spectra, followed
by runs with a modulated carrier to obtain range-Doppler images.  CW
data are typically used to measure total echo power and frequency
extent, whereas range-Doppler images are typically used to resolve the
target in two dimensions.  Table~\ref{tab:observingsummary} summarizes
the CW and range-Doppler imaging runs.

\begin{deluxetable*}{ccccccccccccc}
\tablewidth{0pt} 
\tablecaption{2000~ET70 Radar Observations Log} 
\tablehead{ \colhead{Tel} & \colhead{UT Date} & \colhead{MJD} &
\colhead{Eph} & \colhead{RTT} & \colhead{PTX} &
\colhead{$\delta r$} & 
\colhead{$\delta f$}
& \colhead{$N$} & \colhead{Start-Stop}
& \colhead{Runs} & \colhead{Fig.~\ref{fig:fit_collage}}\\ 
\colhead{} & \colhead{yyyy-mm-dd} & \colhead{} &
\colhead{} & \colhead{(s)} & \colhead{(kW)} &
\colhead{(m)} & \colhead{(Hz)} & \colhead{} & \colhead{hh:mm:ss-hh:mm:ss} & \colhead{} 
& \colhead{key} } 
\startdata
A     & 2012-02-12 & 55969 & s41  &  67 & 828   & cw  &0.167& none  & 08:27:51-08:37:55 & 5       &     \\
  $ $ & $ $        & $ $   & s41  & $ $ &       & 15  &0.075& 65535 & 08:42:47-10:29:47 & 48      & 1-7 \\
  $ $ & $ $        & $ $   & s41  & $ $ &       & 15  &0.075& 8191  & 10:53:18-11:09:55 & 8       &     \\
\\[-2mm]										      									     
\hline\\[-2mm]										      									     
A     & 2012-02-13 & 55970 & s41  &  62 & 860   & cw  &0.182& none  & 08:11:06-08:23:08 & 6       &     \\
  $ $ & $ $        & $ $   & s43  & $ $ &       & 15  &0.075& 65535 & 08:30:34-10:53:26 & 54      & 8-14\\
\\[-2mm]										      									     
\hline\\[-2mm]										      									     
A     & 2012-02-14 & 55971 & s43  &  58 & 811   & cw  &0.196& none  & 07:59:56-08:04:43 & 3       &     \\
  $ $ & $ $        & $ $   & s43  & $ $ &       & 15  &0.075& 65535 & 08:06:40-10:19:45 & 59      &15-22\\
\\[-2mm]										      									     
\hline\\[-2mm]										      									     
A     & 2012-02-15 & 55972 & s43  &  54 & 785   & cw  &0.213& none  & 07:43:02-07:47:29 & 3       &     \\
  $ $ & $ $        & $ $   & s43  & $ $ &       & 15  &0.075& 65535 & 08:03:01-08:09:18 & 4       &     \\
  $ $ & $ $        & $ $   & s47  & $ $ &       & 15  &0.075& 65535 & 08:16:38-10:09:46 & 60      &23-29\\

\\[-2mm]										      									     
\hline\\[-2mm]										      									     
G     & 2012-02-15 & 55972 & s43  &  54 & 420   & cw  &     & none  & 08:55:15-09:03:27 & 5       &     \\
  $ $ & $ $        & $ $   & s43  &     & $ $   & 75  &1.532& 255   & 09:17:48-09:33:20 & 9       &     \\
  $ $ & $ $        & $ $   & s45  & $ $ & $ $   & 37.5&     & 255   & 09:46:24-11:59:24 & 73      &     \\ 
  $ $ & $ $        & $ $   & s45  & $ $ & $ $   & 37.5&     & 255   & 12:15:57-12:24:09 & 5       &     \\  
\\[-2mm]										      									     
\hline\\[-2mm]										      									     
A     & 2012-02-16 & 55973 & s51  &  51 & 760   & cw  &0.227& none  & 07:34:18-07:38:30 & 3       &     \\
  $ $ & $ $        & $ $   & s49  & $ $ &       & 15  &0.075& 65535 & 07:40:56-07:46:52 & 4       &     \\
  $ $ & $ $        & $ $   &  $ $ & $ $ &       & cw  &     & none  & 07:48:38-07:51:06 & 2       &     \\
  $ $ & $ $        & $ $   & s49  & $ $ &       & 15  &0.075& 65535 & 07:53:28-09:38:16 & 62      &30-37\\
\\[-2mm]										      									     
\hline\\[-2mm]										      									     
G     & 2012-02-16 & 55973 & s49  &  51 & 420   & cw  &     & none  & 09:15:46-09:23:30 & 5       &     \\
  $ $ & $ $        & $ $   & s49  &     & $ $   & 75  &1.532& 255   & 09:56:39-10:02:39 & 4       &     \\
  $ $ & $ $        & $ $   & s49  & $ $ & $ $   & 37.5&0.488& none  & 11:35:21-11:46:33 & 7$^*$   &     \\
  $ $ & $ $        & $ $   & s49  & $ $ & $ $   & 15  &1.0  & none  & 12:15:54-12:47:10 & 18$^*$  &43-44\\
  $ $ & $ $        & $ $   & s49  & $ $ & $ $   & 15  &1.0  & none  & 13:10:01-13:28:09 & 11$^*$  &45   \\
  $ $ & $ $        & $ $   & s49  & $ $ & $ $   & 15  &1.0  & none  & 13:29:04-15:29:31 & 70$^*$  &46-53\\
\\[-2mm]										      									     
\hline\\[-2mm]										      									     
A     & 2012-02-17 & 55974 & s53  &  48 & 775   & cw  &0.244& none  & 07:38:00-07:41:57 & 3       &     \\
  $ $ & $ $        & $ $   & s53  & $ $ &       & 15  &0.075& 65535 & 07:44:36-08:48:59 & 40      &38-42\\
\\[-2mm]										      									     
\hline\\[-2mm]										      									     
G     & 2012-02-17 & 55974 & s53  &  48 & 420   & cw  &     & none  & 07:05:53-07:13:10 & 5       &     \\
  $ $ & $ $        & $ $   & s53  &     & $ $   & 75  &1.532& 255   & 07:42:55-08:00:01 & 11      &     \\
  $ $ & $ $        & $ $   & s53  & $ $ & $ $   & 37.5&0.977& none  & 08:16:57-12:24:19 & 152$^*$ &54-65\\
\\[-2mm]										      									     
\hline\\[-2mm]										      									     
G     & 2012-02-18 & 55975 & s53  &  47 & 420   & cw  &     & none  & 07:05:51-07:12:50 & 5       &     \\
  $ $ & $ $        & $ $   & s53  & $ $ & $ $   & 75  &1.532& 255   & 07:36:04-07:50:52 & 10      &     \\
  $ $ & $ $        & $ $   & s53  & $ $ & $ $   & 37.5&0.977& none  & 08:01:15-08:31:44 & 20$^*$  &66   \\
  $ $ & $ $        & $ $   & s53  & $ $ & $ $   & 37.5&0.977& none  & 08:36:51-08:45:24 & 6$^*$   &67   \\
\\[-2mm]										      									     
\hline\\[-2mm]										      									     
G     & 2012-02-19 & 55976 & s55  &  46 & 420   & cw  &     & none  & 07:05:51-07:12:41 & 5       &     \\
  $ $ & $ $        & $ $   & s55  &     & $ $   & 75  &1.532& 255   & 07:21:55-97:36:25 & 10      &     \\
  $ $ & $ $        & $ $   & s55  & $ $ & $ $   & 37.5&0.977& none  & 07:46:12-11:13:58 & 136$^*$ &68-76\\
  $ $ & $ $        & $ $   & s55  & $ $ & $ $   & 37.5&0.977& none  & 11:44:10-13:07:06 & 55$^*$  &77-80\\
\\[-2mm]										      									     
\hline\\[-2mm]										      									     
G     & 2012-02-20 & 55977 & s57  &  46 & 420   & cw  &     & none  & 07:10:51-07:17:42 & 5       &     \\
  $ $ & $ $        & $ $   & s57  &     & $ $   & 37.5&0.977& none  & 08:12:13-09:18:53 & 44$^*$  &81-84\\
  $ $ & $ $        & $ $   & s57  &     & $ $   & 37.5&0.977& none  & 10:31:49-11:33:59 & 41$^*$  &85-87\\
  $ $ & $ $        & $ $   & s57  & $ $ & $ $   & 37.5&0.977& none  & 12:16:08-12:19:54 & 3$^*$   &     \\
\\[-2mm]										      									     
\hline\\[-2mm]										      									     
G     & 2012-02-22 & 55979 & s59  &  48 & 420   & cw  &     & none  & 09:48:25-09:55:42 & 5       &     \\
  $ $ & $ $        & $ $   & s59  &     & $ $   & 75  &0.957& 255   & 10:02:53-10:21:35 & 12      &     \\
  $ $ & $ $        & $ $   & s59  & $ $ & $ $   & 37.5&0.977& none  & 10:32:51-10:35:14 & 2$^*$   &     \\
  $ $ & $ $        & $ $   & s59  & $ $ & $ $   & 37.5&0.977& none  & 10:36:06-10:45:02 & 6$^*$   &     \\
\\[-2mm]										      									     
\hline\\[-2mm]										      									     
G     & 2012-02-23 & 55980 & s59  &  51 & 420   & cw  &     & none  & 08:33:10-08:40:53 & 5       &     \\
  $ $ & $ $        & $ $   & s59  &     & $ $   & 75  &0.957& 255   & 08:49:48-09:09:41 & 12      &     \\
  $ $ & $ $        & $ $   & s59  & $ $ & $ $   & 75  &0.977& none  & 09:20:46-10:55:20 & 55$^*$  &88-90\\
\\[-2mm]										      									     
\hline\\[-2mm]										      									     
A     & 2012-08-24 & 56163 & s72  & 153 & 721   & cw  &     & none  & 15:10:51-15:18:25 & 2       &     \\
      &            &       &      &     &       & cw  &     & none  & 15:46:51-16:31:17 & 9       &     \\
      &            &       &      &     &       & 150  & 0.954   & 8191  & 16:36:59-17:41:01 & 13      &     \\
\\[-2mm]										      									     
\hline\\[-2mm]										      									     
A     & 2012-08-26 & 56165 & s72  & 157 & 722   &cw   &     &  none & 15:04:24-16:20:38 & 15      &     \\
\enddata 

\tablecomments{The first column indicates the telescope: Arecibo
  Planetary Radar (A) or Goldstone Solar System Radar at DSS-14 (G).
  MJD is the modified Julian date of the observation.  Eph is the
  ephemeris solution number used (Section~\ref{sec:orbit}).  RTT is
  the round-trip light-time to the target.  PTX is the transmitter
  power. $\delta r$ and $\delta f$ are the range and Doppler
  resolutions, respectively, of the processed data.  $N$ is the number
  of bauds or the length of the pseudo-random code used.  The timespan
  of the received data is listed by their UT start and stop
  times. Runs is the number of transmit-receive cycles during the
  timespan. An asterisk ($^*$) indicates chirp runs.
  Last column indicates the key to the image numbers shown in
  Fig.~\ref{fig:fit_collage}. }
\label{tab:observingsummary}
\end{deluxetable*}

\begin{figure}
\plotone{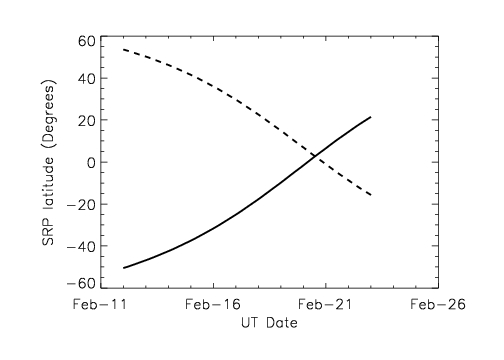}
\caption{
Sub-radar point (SRP) latitude during the primary observing
period. Solid and dashed lines show SRP latitude for the prograde and
retrograde spin vectors, respectively (Section~\ref{sec-spin}).
}

\label{fig:observations}
\end{figure}

For CW runs a carrier wave at a fixed frequency was transmitted for
the RTT to the asteroid.  The received echo from the
asteroid was demodulated, sampled, and recorded. A fast Fourier
transform (FFT) was applied to each echo timeseries to obtain the CW
spectra.
The total frequency extent ($\Delta f$) or bandwidth (BW) of the CW
spectra is equal to the sampling frequency ($f_s$) or the reciprocal
of the sampling period ($P_s$):
\begin{equation}
\Delta f=f_s=\frac{1}{\rm P_s}.
\end{equation}
The spectral resolution ($\delta f$) is given by:
\begin{equation}
\delta f=\frac{\rm \Delta f}{n},
\end{equation}
where $n$ is the FFT length. Finer spectral resolution can be achieved
by increasing $n$, however the signal-to-noise ratio (SNR) in each
frequency bin decreases as ${1}/{\sqrt{n}}$.  The finest possible
resolution that can be achieved is limited by the number of samples
obtained in one run.  If we were recording for the full duration of the
RTT, the number of samples obtained in one run would be $n={\rm RTT}
\times f_s$, and the resolution would be $\delta f~=~1/{\rm RTT}$.
In reality we cannot transmit for a full RTT as it takes several
seconds to switch between transmission and reception, so the finest
possible resolution is $\delta f = 1/({\rm RTT}-t_{\rm switch})$,
where $t_{\rm switch}$ is the switching time.

For range-Doppler imaging two different {\em pulse compression}
techniques were used to achieve fine range resolution while
maintaining adequate SNR~\citep{peebles07}.  Pulse compression is a
signal conditioning and processing technique used in radar systems to
achieve a high range resolution without severely compromising the
ability to detect or image the target.  The range resolution
achievable by a radar is proportional to the pulse duration, or,
equivalently, inversely proportional to the effective bandwidth of the
transmitted signal.
However, decreasing the pulse duration reduces the total transmitted
energy per pulse and hence it negatively affects the ability to detect
the radar target.  Pulse compression techniques allow for the
transmission of a long pulse while still achieving the resolution of a
short pulse.

\subsection{Pulse compression using binary phase coding (BPC)}
We used binary codes to produce range-Doppler images of the asteroid
at both Arecibo and Goldstone. In this technique we modulated the
transmitted carrier with a repeating pseudo-random code using binary
phase shift keying (BPSK).  The code contains $N$ elements or {\em
  bauds} and the duration of each baud is $T_{\rm baud}$.  The
duration of the code is called the pulse repetition period (PRP=$N
\times T_{\rm baud}$).
The effective bandwidth of the transmitted signal ($B_{\rm eff}$) is
given by $1/T_{\rm baud}$.
For each run we transmitted for approximately the RTT to the asteroid,
followed by reception for a similar duration.  The received signal was
demodulated and then decoded by cross-correlating it with a replica of
the transmitted code.  The images span a range ($\Delta r$) given by:
\begin{equation}
\Delta r = \frac{c}{2} {\rm PRP},
\label{eq:totalrange}
\end{equation}
and their range resolution ($\delta r$) is given by:
\begin{equation}
\delta r = \frac{c}{2} T_{\rm baud} = \frac{c}{2}\frac{1}{B_{\rm eff}},
\label{eq:rangeres}
\end{equation}
where $c$ is the speed of light.  Each baud within the code eventually
maps into to a particular range bin in the image.
Resolution in the frequency or Doppler dimension was obtained in each
range bin by performing a FFT on the sequence of returns corresponding
to that bin.  The total frequency extent ($\Delta f$) or bandwidth
(BW) of the image is equal to the pulse repetition frequency (PRF) or the reciprocal of the PRP:
\begin{equation}
\Delta f={\rm PRF}=\frac{1}{\rm PRP}.
\label{eq:bw}
\end{equation}
The frequency resolution ($\delta f$) of the image depends on the
FFT length ($n$) as follows:
\begin{equation}
\delta f = \frac{\Delta f}{n}.
\label{eq:fres}
\end{equation}
The RTT dictates the finest frequency resolution achievable, similar
to the situation with CW spectra.

\subsection{Pulse compression using linear frequency modulation (Chirp)}
We also used a linear frequency modulation
technique~\citep{marg01chirp,peebles07} to produce range-Doppler
images of the asteroid at Goldstone.  Only a fraction of the images
were obtained in this mode because this new capability is still in the
commissioning phase.  Chirp waveforms allow us to maximize the
bandwidth of the transmitted signal and to obtain better range
resolution than that available with BPC waveforms.  They are also less
susceptible to degradation due to the Doppler spread of the targets.
Finally, they are more amenable to the application of windowing
functions that can be used to trade between range resolution and range
sidelobe level~\citep{marg01chirp}.

In this technique the carrier was frequency modulated with a linear
ramp signal. The resultant signal had a frequency that varied linearly
with time from $\omega_0-\omega$ to $\omega_0+\omega$, where
$\omega_0$ is the carrier frequency and $B_{\rm eff}=2\omega$ is the
effective bandwidth of the signal. The resultant signal is called a
chirp. A repeating chirp was transmitted with 100\% duty cycle for the
duration of the round-trip light-time to the asteroid. The time
interval between the transmission of two consecutive chirps is the
PRP. For our observations we used PRPs of 125 $\mu s$ and 50 $\mu s$
for chirps with $B_{\rm eff}$=2~MHz and $B_{\rm eff}$=5~MHz,
respectively.  The received signal was demodulated and range
compression was achieved by cross-correlating the echo with a replica
of the transmitted signal.  The range extent and the range resolution
of the chirp images are given by Eq.~(\ref{eq:totalrange}) and
Eq.~(\ref{eq:rangeres}), respectively.  Resolution in the frequency or
Doppler dimension was obtained as in the binary coding technique.  The
bandwidth of the image is given by Eq.~(\ref{eq:bw}) and the frequency
resolution is given by Eq.~(\ref{eq:fres}).

\section{Astrometry and orbit}
\label{sec:orbit}
A radar astrometric measurement consists of a range or Doppler
estimate of hypothetical echoes from the center of mass (COM) of the
object at a specified coordinated universal time (UTC).  In practice
we used our measurements of the position of the leading edge of the
echoes and our estimates of the object's size to report preliminary
COM range estimates and uncertainties.  We refined those estimates
after we obtained a detailed shape model (Section~\ref{sec-shape}).
We measured Doppler astrometry from the CW spectra.  We reported 9
range estimates and 1 Doppler estimate during the course of the
observing run and computed the heliocentric orbit using the JPL
on-site orbit determination software (OSOD).  The ephemeris solution
was updated each time new astrometric measurements were incorporated
(Table~\ref{tab:observingsummary}).  Using the best ephemeris solution
at any given time minimizes smearing of the images.

At the end of our February observing campaign, we were using ephemeris
solution 59. A final range measurement, obtained during the asteroid's
close approach in August 2012, was incorporated to generate orbit
solution 74. After the shape model was finalized, we updated the orbit
to solution 76 by replacing the preliminary February astrometric
measurements with more accurate shape-based estimates of the range to
the center of mass (Table~\ref{tab:astrometry}).

Table~\ref{tab:orbit} lists the best fit orbital parameters (solution
76) generated using 18 range measurements and 316 optical
measurements. The optical measurements span February 1977 to December
2012.  However we assigned 20 arcsecond uncertainties to the two
precovery observations from the 1977 La Silla-DSS plates, effectively
removing their contribution to the fit and reducing the optical arc to
the interval 2000-2012.  The 1977 observations were from a single,
hour-long, trailed exposure, and appear to have been reported with a
$\sim$45~s timing error, consistent with the measurers' cautionary
note ``Start time and exposure length are uncertain'' (MPEC 2000-L19).

The orbit computation is reliable over a period from the year 460 to
2813. Beyond this interval either the $3\sigma$ uncertainty of the
Earth close approach time (evaluated whenever the close approach
distance is less than 0.1 AU) exceeds 10 days or the $3\sigma$
uncertainty of the Earth close approach distance exceeds 0.1 AU.
The current {\em Minimum Orbit Intersection Distance} (MOID) with
respect to Earth is 0.03154 AU, making 2000~ET70 a potentially
hazardous asteroid (PHA).

\begin{deluxetable*}{lrcc}
\tablecaption{2000~ET70 range measurements} 
\tablehead{\colhead{Date (UTC)}  &
  \colhead{Range} & \colhead{1-$\sigma$ Uncertainty} & \colhead{Observatory} \\
  \colhead{yyyy-mm-dd hh:mm:ss} & \colhead{$\mu s$} &
  \colhead{$\mu s$} }
\startdata

2012-02-12 08:50:00 &	67220894.76  &	0.5   & A	\\
2012-02-12 10:03:00 &	66974197.50  &	0.5   & A	\\	
2012-02-13 09:11:00 &	62452848.06  &	0.5   & A	\\	
2012-02-13 10:00:00 &	62298918.42  &	0.5   & A	\\	
2012-02-14 09:28:00 &	58061903.74  &	0.5   & A	\\	
2012-02-14 10:06:00 &	57953990.18  &	0.5   & A	\\	
2012-02-14 10:16:00 &	57925824.25  &	0.5   & A	\\	
2012-02-15 08:30:00 &	54319877.92  &	0.5   & A	\\	
2012-02-15 09:20:00 &	54206358.81  &	3.0   & G	\\	
2012-02-15 09:47:00 &	54126135.16  &	0.5   & A	\\	
2012-02-16 08:23:00 &	50979848.61  &	0.5   & A	\\	
2012-02-16 09:30:00 &	50840187.02  &	0.5   & A	\\	
2012-02-17 08:05:00 &	48327621.88  &	0.5   & A	\\	
2012-02-17 08:43:00 &	48267222.55  &	0.5   & A	\\	
2012-02-18 07:40:00 &	46481930.28  &	2.0   & G	\\
2012-02-19 07:30:00 &	45481292.56  &	2.0   & G	\\
2012-02-20 10:10:00 &	45483470.81  &	2.0   & G	\\
2012-08-24 17:08:00 &	153139162.28 &	3.0   & A	\\ 
\enddata 

\tablecomments{This table lists shape-based estimates of the range to
  the asteroid COM.  The first column indicates the coordinated
  universal time (UTC) of the measurement epoch.  The second column
  gives the ranges expressed as the RTT to the asteroid in
  microseconds ($\mu s$).  The third column lists the 1$\sigma$ range
  uncertainty.  The fourth column indicates the radar used to make the
  measurement (A stands for the Arecibo Planetary Radar and G stands
  for the Goldstone Solar System Radar at DSS-14).}
\label{tab:astrometry}
\end{deluxetable*}

\begin{deluxetable}{lrr}
\tablecaption{2000~ET70 heliocentric orbital elements (solution 76)}
\tablehead{\colhead{Element} & \colhead{Value} & \colhead{1-$\sigma$ Uncertainty}\\ 
\colhead{} & \colhead{} &\colhead{}}
\startdata
eccentricity             & 0.123620379       & $6.3 \times 10^{-8}$ \\ 
\\
semi-major axis (AU)     & 0.9466347364      & $1.2 \times 10^{-9}$ \\
\\
inclination (degrees)    & 22.3232174     & $1.2 \times 10^{-6}$ \\
\\
longitude of \\
ascending node (degrees) & 331.16730395     & $9.8 \times 10^{-7}$ \\
\\
argument of \\
perihelion (degrees)     & 46.106698     & $1.1 \times 10^{-5}$ \\
\\
Mean anomaly (degrees)   & 84.37370818     & $4.2 \times 10^{-7}$ \\
\enddata
\tablecomments{All orbital elements are specified at epoch 
2012 Dec 15.0 
barycentric dynamical time (TDB) in the heliocentric ecliptic
reference frame of J2000.  The corresponding orbital period is
(336.41246710~$\pm~5.5 \times 10^{-7}$) days.}
\label{tab:orbit}
\end{deluxetable}

\section{Radar scattering properties}
\label{sec:radar_scatter}
\begin{figure*}
\epsscale{0.85}
\plottwo{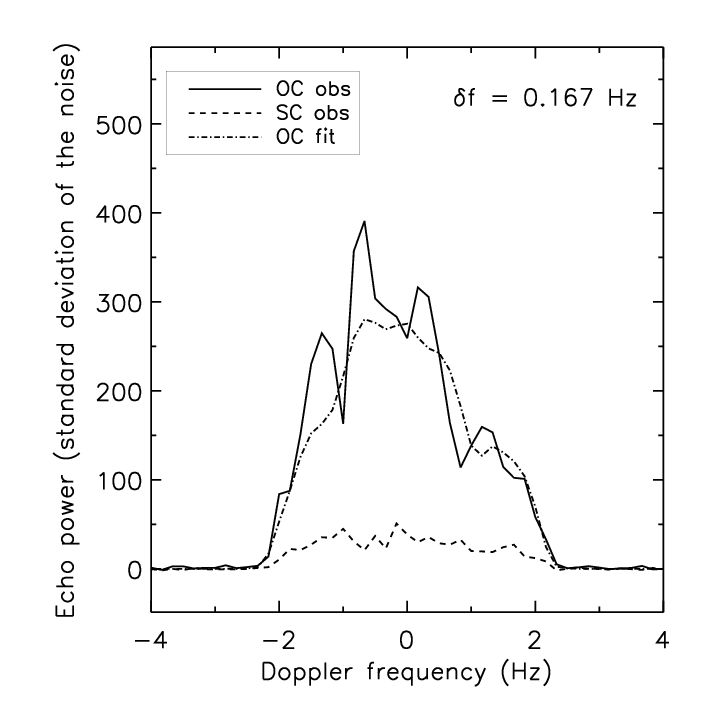}{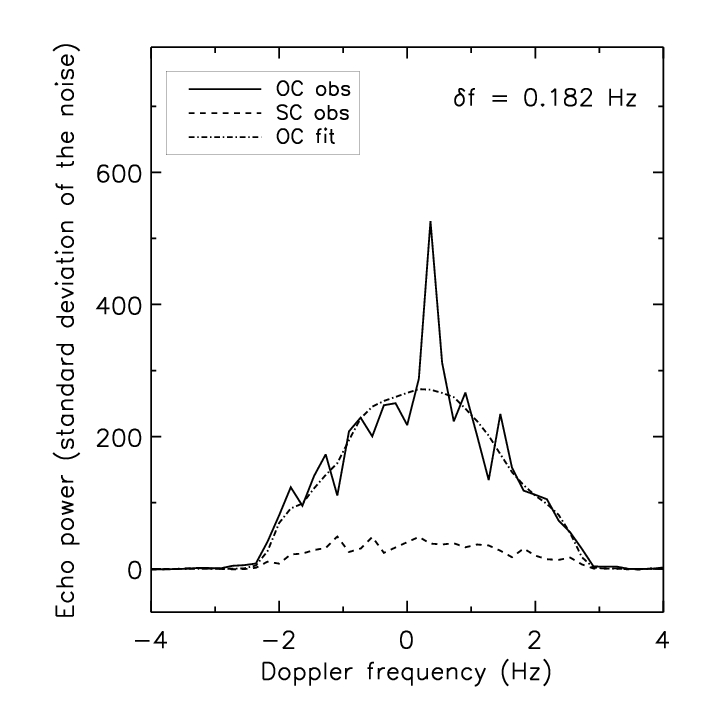}
\plottwo{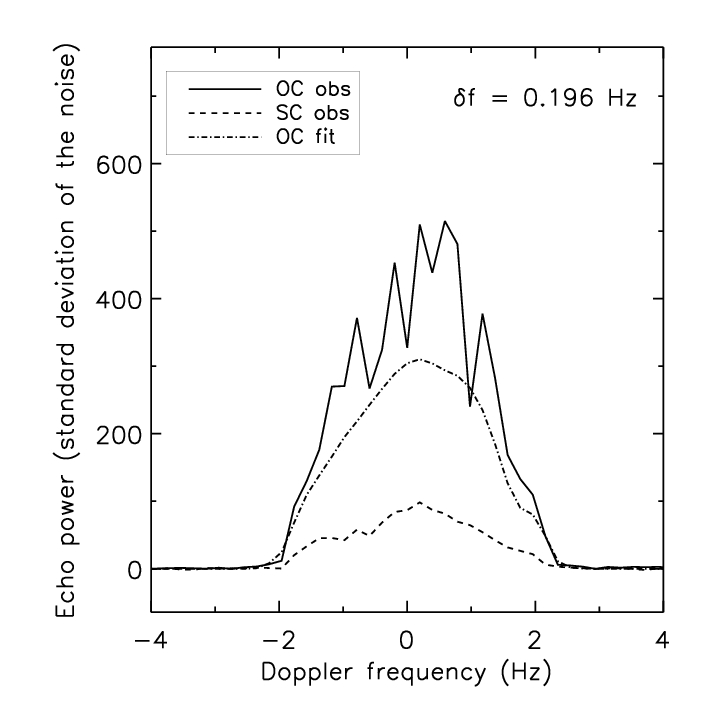}{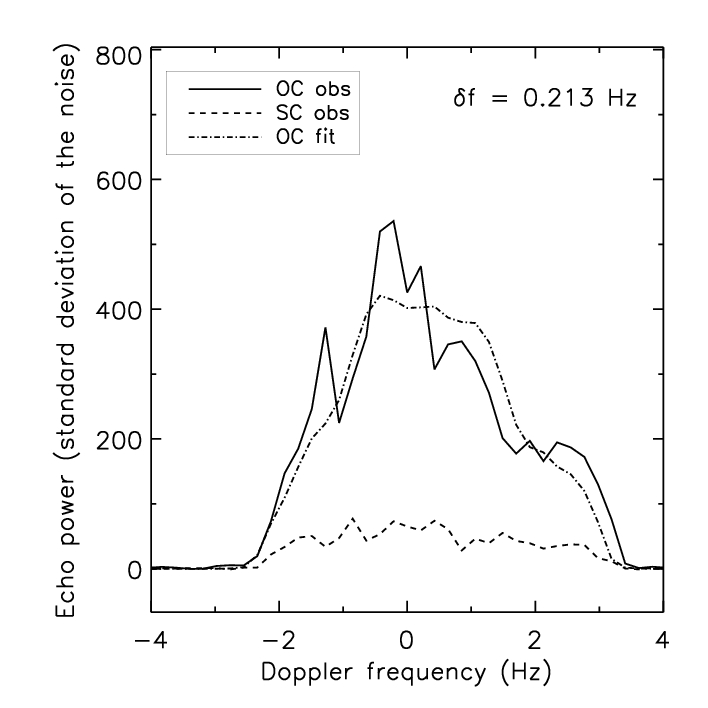}
\plottwo{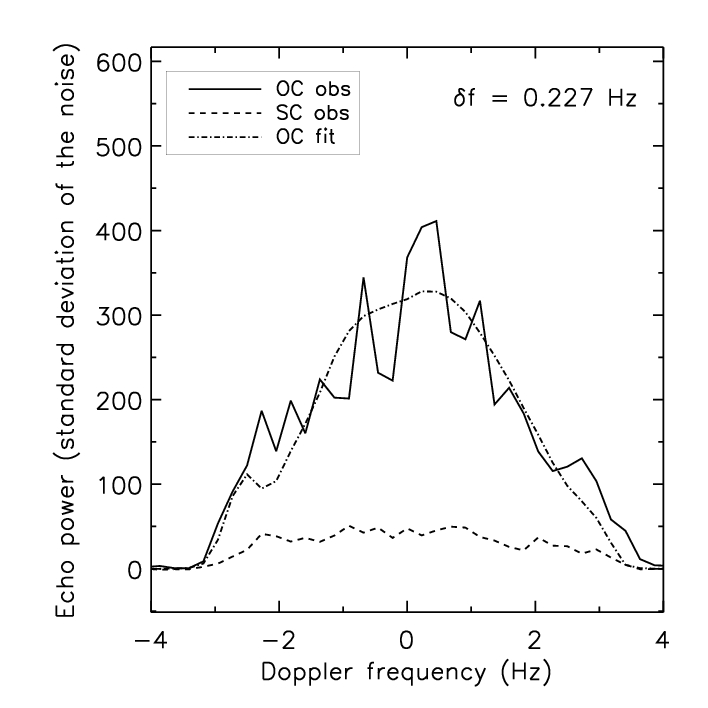}{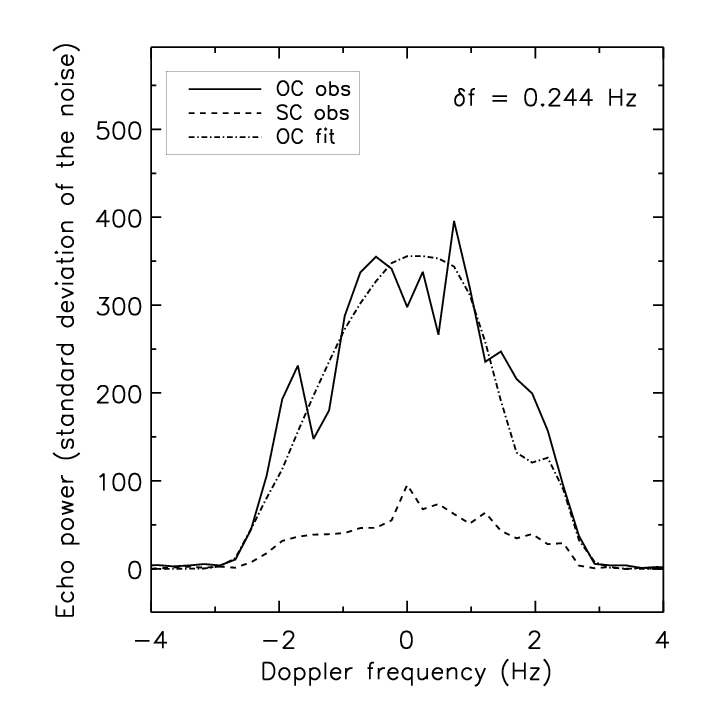}
\caption{Arecibo CW spectra of 2000~ET70 arranged in chronological
  order from left to right and top to bottom.  They were obtained 
  on February 12-17 (MJDs 55969.35, 55970.34, 55971.33, 55972.32,
  55973.32, 55974.32, respectively).  The frequency resolution of each
  spectrum is given at the top right of the corresponding panel.  Each
  spectrum was produced by choosing a frequency resolution that
  allowed for the incoherent sum of 10 independent spectra, or {\em
    looks}, per run, and by summing over 3 runs, resulting in a total
  number of 30 looks.
The solid and dashed lines are observed OC and SC spectra,
respectively. The dot-dashed line shows the corresponding synthetic OC
spectra generated using our shape model (Section~\ref{sec-shape}).
The circular polarization ratio ($\mu_{\rm C}$) values in
chronological order are 0.18, 0.20, 0.24, 0.21, 0.21, and 0.24 all of
which have uncertainties of 5\%, where the uncertainty is computed
according to \citet{ostro83}. 
The OC radar albedo values in chronological order are 0.091, 0.063,
0.073, 0.056, 0.051, and 0.044 all of which have uncertainties of 25\%.
The reduced chi-squares of the fits to the OC spectra vary between 0.69 to 0.72.  }

\label{fig:cw}
\end{figure*}

We transmitted circularly polarized waves and used two separate
channels to receive echoes having the same circular (SC) and the
opposite circular (OC) polarization as that of the transmitted
wave~\citep{ostro93}. Reflections from a plane surface reverse the
polarization of the incident waves and most of the echo power is
expected in the OC polarization. Echo power in the SC polarization is
due to multiple reflections or reflections from structures with
wavelength-scale roughness at the surface or sub-surface.  A higher
ratio of SC to OC power therefore indicates a greater degree of
near-surface wavelength-scale roughness or multiple scattering.  It is
worth noting that while a larger SC to OC ratio implies a rougher
surface, there is a compositional component at well~\citep{benner08}.
This {\em circular polarization ratio} is often denoted by $\mu_{\rm
C}$.  We measured $\mu_{\rm C}$ for all the Arecibo spectra shown in
Fig.~\ref{fig:cw} and computed an average value of $\mu_{\rm C} =
0.21~\pm~0.02$, where the uncertainty is the standard deviation of the
individual estimates.  Observed ratios for individual spectra deviate
no more than 0.03 from the average.  This ratio is lower than that for
the majority of NEAs with known circular polarization ratios (Mean =
0.34~$\pm$~0.25, Median = 0.26)~\citep{benner08} suggesting that
2000~ET70 has a lower than average near-surface roughness at 10 cm
scales.

The average radar albedo computed for the OC CW spectra shown in
Fig.~\ref{fig:cw} is $0.063~\pm~0.017$, where the uncertainty is the
standard deviation of individual estimates.  The radar albedo is the
ratio of the radar cross-section to the geometric cross-sectional area
of the target. The radar cross-section is the projected area of a
perfectly reflective isotropic scatterer that would return the same
power at the receiver as the target.

In our modeling of the shape of the asteroid (Sections~\ref{sec-spin}
and \ref{sec-shape}), we used a cosine law to represent the radar
scattering properties of 2000~ET70:
\begin{equation}
\frac{d\sigma}{dA}=R(C+1)(\cos{\alpha})^{2C}. 
\label{eq-scattering}
\end{equation}
Here $\sigma$ is the radar cross section, $A$ is the target surface
area, $R$ is the Fresnel reflectivity, $C$ is a parameter describing
the wavelength-scale roughness, and $\alpha$ is the incidence angle.
Values of $C$ close to 1 represent diffuse scattering, whereas larger
values represent more specular scattering~\citep{mitchell96}.
For specular scattering, $C$ is related to the
wavelength-scale adirectional root-mean-square (RMS) slope $S_0$ and
angle $\theta_{\rm rms}$ of the surface by $S_0=\tan{(\theta_{\rm
rms})}=C^{-1/2}$.

\section{Range and Doppler extents}
The range extent of the object in the radar images varies between
$\sim$600~m
and $\sim$1700~m (Fig.~\ref{fig:fit_collage}),
suggesting that the asteroid is significantly elongated. In most of
the images two distinct ridges that surround a concavity are clearly
visible (e.g., Fig.~\ref{fig:fit_collage}, images 8-12 and images
30-37).  In images where the ridges are aligned with the Doppler axis,
they span almost the entire bandwidth extent of the asteroid (e.g.,
Fig.~\ref{fig:fit_collage}, images 11 and 34). If the concavity is a
crater then these ridges could mark its rim.  At particular viewing
geometries, the trailing end of the asteroid exhibits a large outcrop
with a range extent of $\sim$250~m (e.g., Fig.~\ref{fig:fit_collage},
image 11).  These features suggest that the overall surface of
the asteroid is highly irregular at scales of hundreds of meters.

For a spherical object, the bandwidth ($B$) of the radar echo is given
by:
\begin{equation}
B=\frac{4 \pi D}{\lambda P}\cos\delta.
\label{eq:bw_srplat}
\end{equation}
Here $D$ is the diameter of the object, $P$ is its apparent spin
period, $\lambda$ is the radar wavelength, and $\delta$ is the
sub-radar latitude.  As $\delta$ increases, $B$ decreases. In images
obtained at similar rotational phases, the bandwidth extent of the
asteroid increased from February 12 to 20, indicating that our view
was more equatorial towards the end of the observing campaign
(Fig.~\ref{fig:observations}).  For example in
Fig.~\ref{fig:fit_collage}, images 10, 33, and 68 are at similar
rotational phases and their bandwidths (based on a 2380 MHz carrier)
are $\sim$3.7~Hz, $\sim$4.9~Hz, and $\sim$6.3~Hz, respectively.

\section{Spin vector}
\label{sec-spin}
We used the SHAPE software~\citep{hudson93,magri07} to fit a shape
model to the radar images and to estimate the spin vector of
2000~ET70. Since SHAPE is not particularly effective at fitting the
spin axis orientation and spin period of the shape model, we carried
out an extensive search for these parameters in an iterative manner.
We performed two iterations each in our search for the spin axis
orientation and spin period, where the result of each step provides
initial conditions for the next optimization step.  This approach
leads to increased confidence that a global minimum is reached.

\begin{figure}
\begin{center}

\plotone{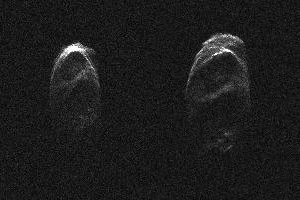}\\
\plotone{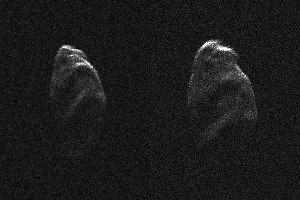}\\
\plotone{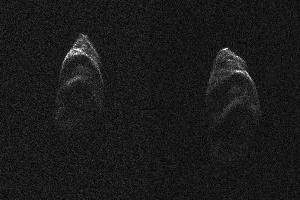}

\end{center}

\caption{2000~ET70 range-Doppler images showing the asteroid at
  similar rotational phases on different days.  The images were
  obtained at MJDs 55970.40 and 55973.38 ($\sim$72 hours apart) (top),
  55971.35 and 55974.33 ($\sim$72 hours apart) (center), 55969.46 and
  55971.34 ($\sim$45 hours apart) (bottom), suggesting a spin period
  of $\sim$9 hours or a sub-multiple of $\sim$9 hours.  In these
  images radar illumination is from the top, range increases towards
  the bottom, Doppler frequency increases to the right, and the
  asteroid spin results in counter-clockwise rotation.}
\label{fig:rp_match}
\end{figure}

\begin{figure}
\begin{center}

\plotone{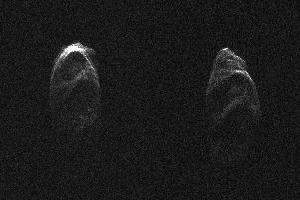}

\end{center}
\caption{2000~ET70 images obtained at MJDs 55970.40 and 55971.34, or
  22.5 hours apart. The asteroid is not close to similar rotational
  phases in these images, ruling out spin periods that are
  sub-multiples of $\sim$22.5 hours.  A $\sim$4.5-hour spin period is
  therefore ruled out, leaving $\sim$9 hours as the only plausible
  value.}
\label{fig:rp_mismatch}
\end{figure}

Our initial estimate of the spin period came from the time interval
between repeating rotational phases of the object captured in the
images.  Fig.~\ref{fig:rp_match} shows the object at similar
orientations in images taken on different days. The time interval in
two of those cases is $\sim$72 hours and in the third case is $\sim$45
hours.  This indicates that the spin period of the object is close to
a common factor of the two, that is, 9 hours or a sub-multiple of 9
hours.  Fig.~\ref{fig:rp_mismatch} shows two images taken ~22.5 hours
apart and the object is not close to similar orientations in these two
images, ruling out all periods that are factors of 22.5 hours.  Thus
we are left with a period close to 
3 hours or 9 hours. Images obtained over observing runs longer
than 3 hours do not show a full rotation of the asteroid, ruling out a
spin period of 3 hours.

Using an initially fixed spin period of 9 hours, we performed an
extensive search for the spin axis orientation. The search consisted
of fitting shape models to the images under various assumptions for
the spin axis orientation.  We covered the entire celestial sphere
with uniform angular separations of 15$^\circ$ between neighboring
trial poles.
At this stage we used a subset of the images in order to decrease the
computational burden.  We chose images showing sharp features that
repeated on different days because they provide good constraints on
the spin state.  Images obtained on February 12 and 15 satisfied these
criteria and we used images with receive times spanning MJDs 55969.426
to 55969.463 and 55972.335 to 55972.423 for this search.  We started
with triaxial ellipsoid shapes and allowed the least-squares fitting
procedure to adjust all three ellipsoid dimensions in order to provide
the best match between model and images.  We then used shapes defined
by a vertex model with 500 vertices and 996 triangular facets.  The
fitting procedure was allowed to adjust the positions of the vertices
to minimize the misfit.  In all of these fits the spin axis
orientation and the spin rate were held constant.  Radar scattering
parameters $R$ and $C$ described in Eq.~(\ref{eq-scattering})
were allowed to float.  We found the best shape model fit with the
spin pole at ecliptic longitude ($\lambda$) = 60$^\circ$ and ecliptic
latitude ($\beta$) = -60$^\circ$.

Using the best fit spin pole from the previous step, we proceeded to
estimate the spin period with greater precision. We tried spin rates
in increments of 2$^\circ$/day from 960$^\circ$/day (P = 9 hours)
to 970$^\circ$/day (P = 8.91 hours) to fit vertex shape models
with 500 vertices to the images. This time we used a more extensive
dataset consisting of all images obtained from Arecibo. As in the
previous step, only the radar scattering parameters and the shape
parameters were allowed to float in addition to the parameter of
interest. We found the best agreement between model and observations
with a spin rate of 964$^\circ$/day (Period = 8.963 hours).

\begin{figure*}
\plotone{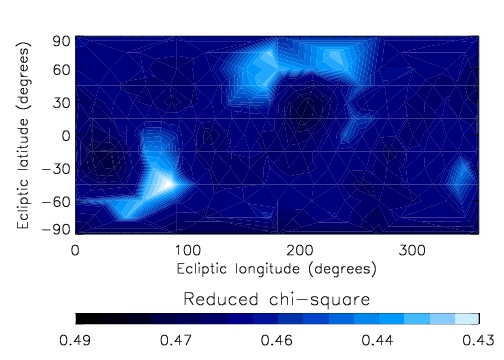}
\caption{Contour plot showing goodness of fit (reduced chi-square) of
shape model fits under various assumptions for the ecliptic longitude
($\lambda$) and ecliptic latitude ($\beta$) of the spin axis. 
The spin period was fixed at 8.963~hours for the fits.
We adopted a best fit spin axis orientation of $\lambda=80^\circ$ and
$\beta=-50^\circ$ with a 10$^\circ$ uncertainty.}
\label{fig:gridsearch}
\end{figure*}

The second iteration of the spin-axis orientation search was similar
to the first one except that we used a spin period of 8.963 hours and
used the complete dataset consisting of all the Arecibo and Goldstone
images from February.  The Arecibo images from August were not used
because of their low SNR and resolution.  This procedure was very
effective in constraining the possible spin axis orientations to a
small region (around $\lambda=79^\circ$ and $\beta=-42^\circ$) of the
celestial sphere (Fig.~\ref{fig:gridsearch}).
We performed a higher resolution
search within this region with spin poles ranging in $\lambda$ from
64$^\circ$ to 104$^\circ$ and $\beta$ from -60$^\circ$ to -30$^\circ$
with step sizes of 4$^\circ$ in $\lambda$ and 5$^\circ$ in $\beta$.
For this step we performed triaxial ellipsoid fits followed by
spherical harmonics model fits, adjusting spherical harmonic
coefficients up to degree and order 10.  Our best estimate of the spin
axis orientation is $\lambda=80^\circ$ and $\beta=-50^\circ$, with
10$^\circ$ uncertainties.  Shape models with spin axis orientations
within this region have similar appearance upon visual inspection.
We also attempted shape model fits with the best-fit prograde pole at
$\lambda=232^\circ$ and $\beta=75^\circ$.  We found that we were
unable to match the observed bandwidths and ruled out the prograde solution.
2000~ET70 is a retrograde spinner, just like the majority of
NEAs~\citep{spina04}.  Our adopted spin pole ($\lambda=80^\circ$,
$\beta=-50^\circ$) is at an angle of $\sim$160$^\circ$ from the
heliocentric orbit pole ($\lambda=241^\circ$, $\beta=68^\circ$).

\begin{figure}
\plotone{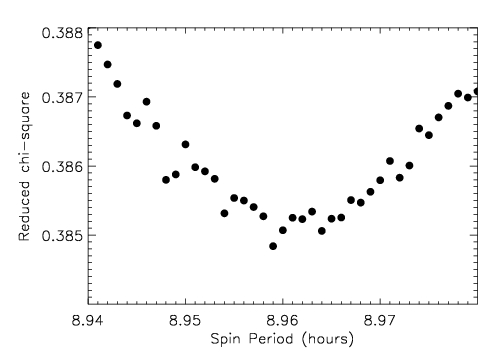}
\caption{Goodness of fit (reduced chi-square) of degree-and-order 10
spherical harmonics shape models of 2000~ET70 as a function of assumed
spin period. 
The spin pole was fixed at $\lambda=80^\circ$ and $\beta=-50^\circ$
for the fits.  We adopted a best fit spin period of 8.960$\pm$0.01
hours.}
\label{fig:spinrate}
\end{figure}

We used the best spherical harmonics shape model from the previous
search to perform a second iteration of the spin period search. This
time we fit spherical harmonics shape models using spin periods in
increments of 0.001 hours from 8.940 hours to 8.980 hours.  This final
step was performed in part to quantify error bars on the spin period.
The reduced chi-squares of the shape models, computed according to the
method described in~\citep{magri07}, are shown in
Fig.~\ref{fig:spinrate}.  We visually verified the quality of the fits
and adopted a spin period of 8.960~$\pm$~0.01 hours.  A 0.01 hour
difference in spin period amounts to a $\sim$13$^\circ$ offset in
rotational phase over the primary observing window which is
detectable.

\section{Shape model}
\label{sec-shape}

\begin{figure*}
\plotone{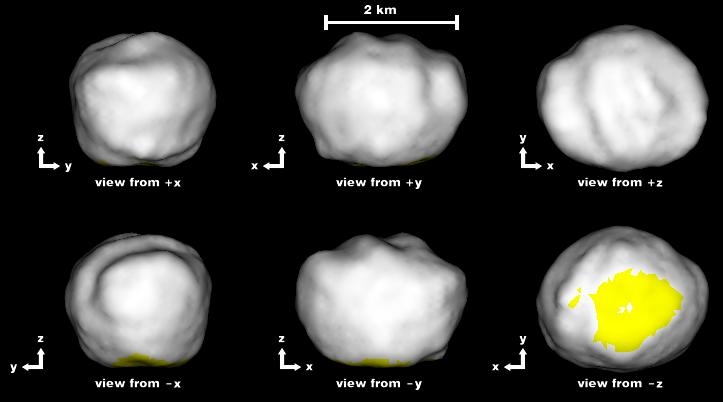}
\caption{Best-fit vertex shape model of 2000~ET70 as seen along the
three principal axes x, y, and z. For principal axis rotation the spin
axis is aligned with the z axis. Yellow regions at the south pole
have radar incidence angles $>60^\circ$ and hence are not well
constrained. The shape model has 2000 vertices and 3996 triangular
facets giving an effective surface resolution of $\sim$100~m.}
\label{fig:shape}
\end{figure*}

\begin{figure*}
\includegraphics[width=7.1in, height=9.2in]{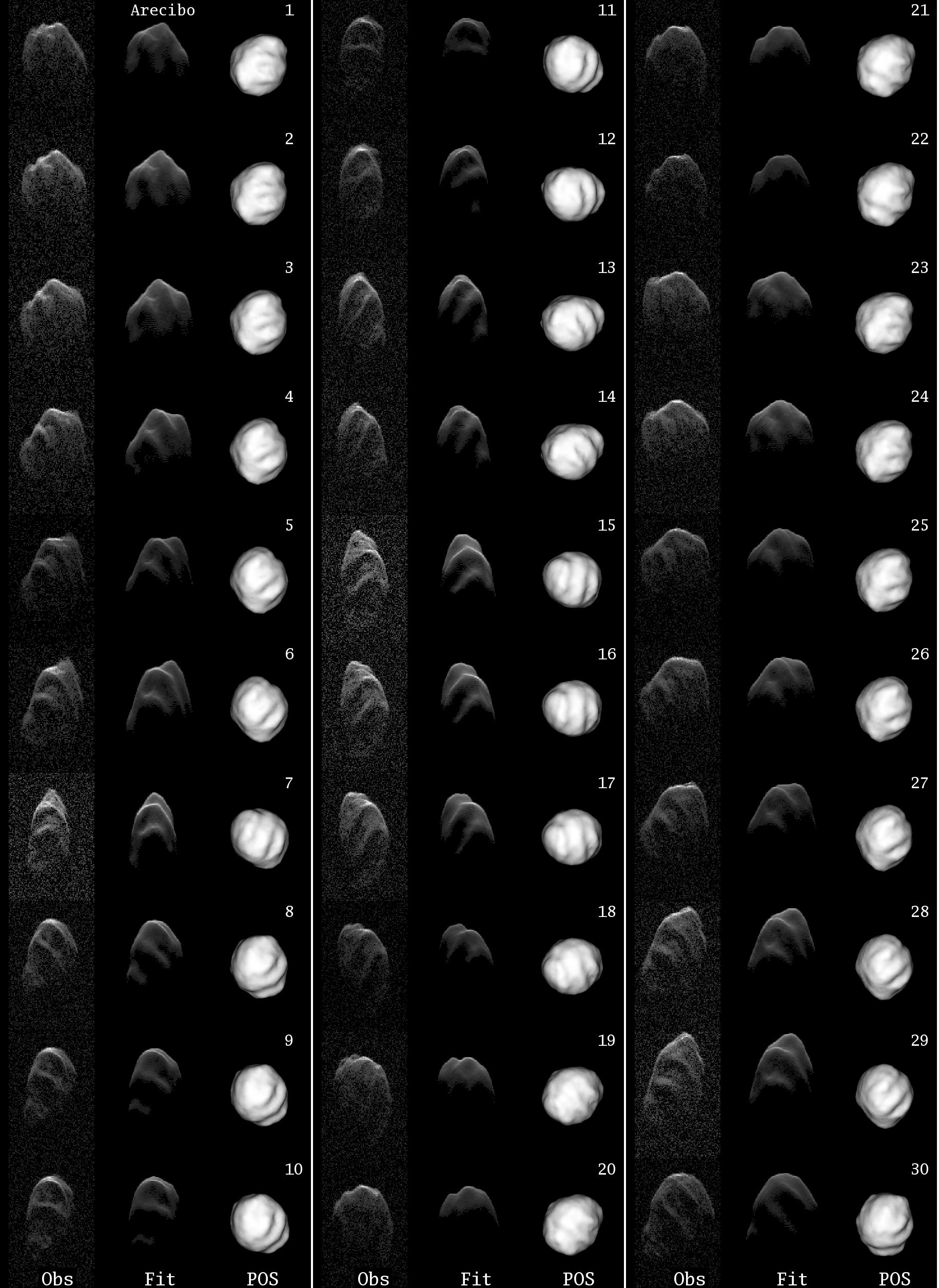}

\caption{Comparison between radar range-Doppler images, corresponding
  synthetic radar images generated using our best-fit shape model, and
  plane-of-sky (POS) projections of the shape model.  Range-Doppler
  images are oriented such that radar illumination is from the top,
  range increases towards the bottom, Doppler frequency increases to
  the right, and the asteroid spin results in counter-clockwise
  rotation. The POS projections are oriented north-up and east-left.
  Time increases from top to bottom within each panel and from left to
  right.  The range and frequency resolutions of the images are given
  in Table~\ref{tab:observingsummary}.  The reduced chi-square of the
  fit to the images is 0.27.
}

\label{fig:fit_collage}
\end{figure*}

As a starting point for our final shape modeling efforts we used the
results of the spin state determination (Section~\ref{sec-spin}),
specifically the best-fit spherical harmonics shape model with a spin
period of 8.960 hours and a spin pole at $\lambda=80^\circ$ and
$\beta=-50^\circ$.  We proceeded to fit all the radar images obtained
in February and OC CW spectra from Fig.~\ref{fig:cw} with a vertex
model having 2000 vertices and 3996 facets.  The number of vertices is
based on experience and the desire to reproduce detectable features
without over-interpretation.  At this step the vertex locations and
the radar scattering parameters were fit for, but the spin vector was
held fixed.  Observed images were summed to improve their SNR. At
Arecibo, 8 images were typically combined. At Goldstone, 14, 9, 13,
13, 14, 12, 6, and 12 images were typically combined on February 15,
16, 17, 18, 19, 20, 22, and 23, respectively.  We cropped the images so
that sufficient sky background remained for the computation of noise
statistics, but the optimization procedure is robust against the
amount of sky background.  We minimized an objective function that
consists of the sum of squares of residuals between model and actual
images, plus a number of weighted penalty functions designed to favor
models with uniform density, principal axis rotation, and a reasonably
smooth surface~\citep{hudson93,magri07}.
The choice of weights in the
penalty function is subjective, so the shape model solution is not
unique. We tried to restrict the weights to the minimum value at which
the penalty functions were effective.  The minimization procedure with
our choice of penalty functions produced a detailed shape model for
2000~ET70 (Fig.~\ref{fig:shape} and Table~\ref{tab:shapemodel}).  The
agreement between model and data is generally excellent but minor
disagreements are observed (Fig.~\ref{fig:fit_collage} and
Fig.~\ref{fig:cw}).  The overall shape is roughly a triaxial ellipsoid
with extents along the principal axes of $\sim$2.61~km, $\sim$2.22~km,
and $\sim$2.04~km, which are roughly the same as the {\em dynamically
equivalent equal volume ellipsoid} (DEEVE) dimensions listed in
table~\ref{tab:shapemodel}.

The region around the north pole has two ridges that are 1-1.5~km in
length and almost 100~m higher than their surroundings.  These ridges
enclose a concavity that seems more asymmetric than most impact
craters.  Along the negative x-axis a large protrusion is visible.
Such a feature could arise if the asteroid were made up of multiple
large components resting on each other.

NEAs in this size range for which radar shape models exist commonly
exhibit irregular features such as concavities and ridges.  A few
examples include Golevka~\citep{hudson00}, 1992 SK~\citep{busch06},
and 1998 WT24~\citep{busch08}.  
The concavities observed on these NEAs 
can
not be adequately captured
by convex-only shape modeling techniques.

For this shape model, the best fit values for the radar scattering
parameters, $R$ and $C$, were 1.9 and 1.4,
respectively~(section~\ref{sec:radar_scatter}).
This value of $C$ indicates that 2000~ET70 is a diffuse scatterer, simlar to
other NEAs such as Geographos~\citep{hudson99}, Golevka~\citep{hudson00},
and 1998~ML14~\citep{ostro01}.  
NEAs are generally expected to be diffuse scatterers at radar
wavelengths because of their small sizes and rough surfaces.  Attempts
to fit the echoes with a two-component scattering law (diffuse plus
specular) did not yield significantly better results.

\begin{figure*}
\includegraphics[width=7.1in, height=9.2in]{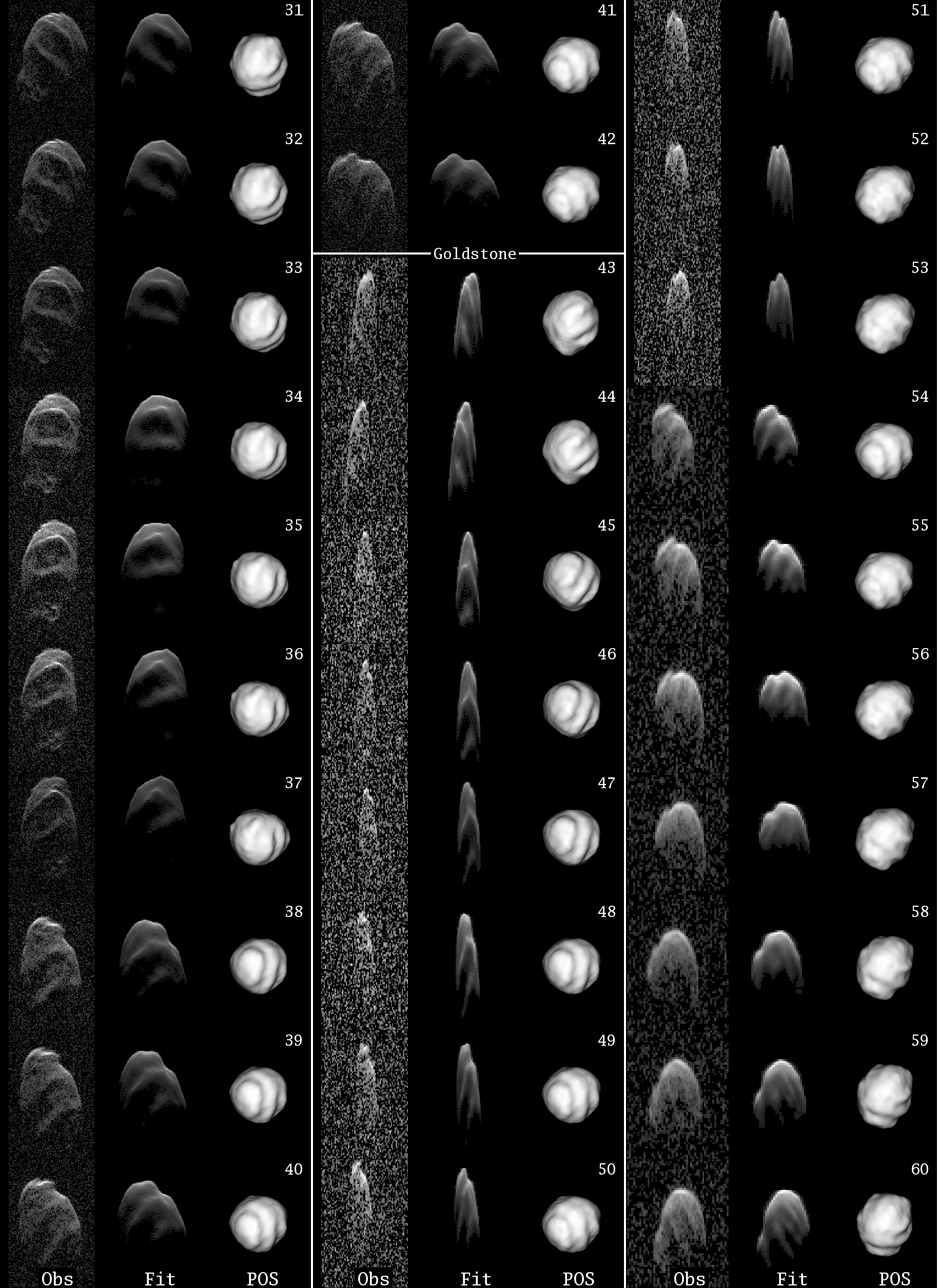}
\end{figure*}

\begin{figure*}
\includegraphics[width=7.1in, height=9.2in]{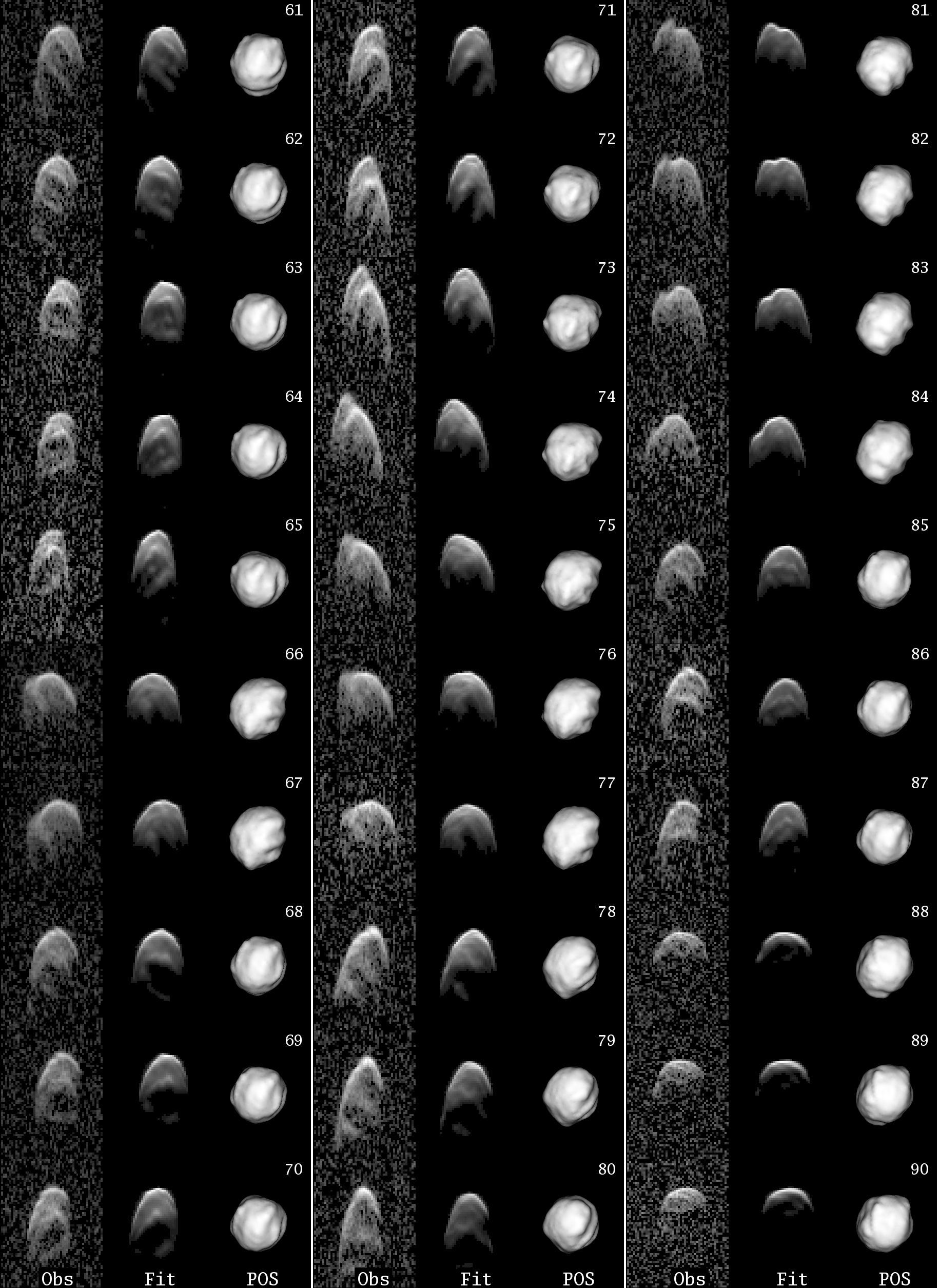}

\end{figure*}

\begin{deluxetable}{lccc}
\tablewidth{0pt}
\tablecaption{2000~ET70 shape model parameters}
\tablehead{\colhead{Parameters} & \colhead{} & \colhead{Value}}
\startdata

Extents along                     &  x        & 2.61 $\pm$ 5\% \\
principal axes (km)               &  y        & 2.22 $\pm$ 5\% \\
                                  &  z        & 2.04 $\pm$ 5\% \\
\\
Surface Area (km$^2$)             &           & 16.7 $\pm$ 10\% \\
\\
Volume (km$^3$)                   &           & 6.07 $\pm$ 15\% \\
\\
Moment of inertia ratios          & $A/C$     & 0.800 $\pm$ 10\% \\
                                  & $B/C$     & 0.956 $\pm$ 10\% \\
\\
Equivalent diameter (km)          &           & 2.26 $\pm$ 5\% \\
\\ 
DEEVE extents (km)                & x         & 2.56 $\pm$ 5\% \\ 
                                  & y         & 2.19 $\pm$ 5\% \\ 
                                  & z         & 2.07 $\pm$ 5\% \\ 
\\
Spin pole ($\lambda, \beta$) ($^\circ$) &     & (80, -50) $\pm$ 10 \\
Sidereal spin period (hours)      &           & 8.96 $\pm$ 0.01 \\
\enddata 

\tablecomments{The shape model consists of 2000 vertices and
3996 triangular facets, corresponding to an effective surface
resolution of $\sim$100 m. The moment of inertia ratios were
calculated assuming homogeneous density.  Here $A$, $B$, and $C$ are
the principal moments of inertia, such that $A<B<C$.  Equivalent
diameter is the diameter of a sphere having the same volume as that of
the shape model.  A {\em dynamically equivalent equal volume
ellipsoid} (DEEVE) is an ellipsoid with uniform density having the
same volume and moment of inertia ratios as the shape model. 
}
\label{tab:shapemodel}
\end{deluxetable}

\section{Gravitational environment}
\begin{figure*}
\plotone{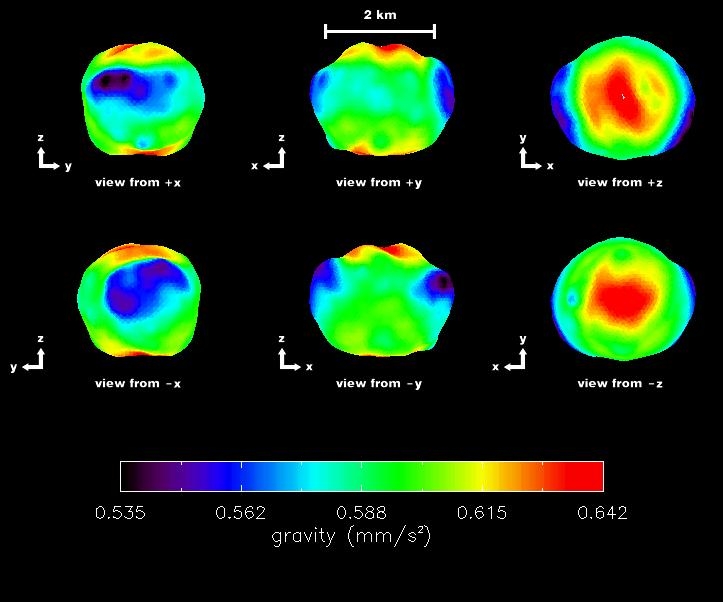}
\caption{This figure shows the magnitude of the vector sum of
acceleration due to gravity and centrifugal acceleration computed at
the center of each facet of the shape model.  We assumed a uniform
density of 2000~kg~m$^{-3}$ and used our measured spin period value of
8.96 hours. Centrifugal acceleration makes a significant contribution
to the total acceleration; at the most protruding regions of the
equator it accounts for about 10\%.  }
\label{fig:gravity_map}
\end{figure*}

We used our best-fit shape model and a uniform density assumption of
2000~kg~m$^{-3}$, a reasonable density for rubble-pile NEAs,
~\citep[e.g.,][]{ostro06}, to compute the gravity field at the surface
of the asteroid~\citep{werner97}.
The acceleration on the surface is
the sum of the gravitational acceleration due to the asteroid's mass
and the centrifugal acceleration due to the asteroid's spin. An
acceleration vector was computed at the center of each
facet. Figure~\ref{fig:gravity_map} shows the variation of the magnitude
of this acceleration over the surface of the asteroid. The
acceleration on the surface varies between 0.54~mm~s$^{-2}$ to
0.64~mm~s$^{-2}$, which is 4 orders of magnitude smaller than that
experienced on Earth and 2 orders of magnitude smaller than that on
Vesta.  Centrifugal acceleration makes a significant contribution to
the total acceleration and varies from zero at the poles to
$\sim$0.049~mm~s$^{-2}$ (about 10\% of the total acceleration) on the
most protruding regions of the equator. For comparison, on Earth,
centrifugal acceleration contributes less than 0.5\% to the total
acceleration at the equator.

\begin{figure*}
\plotone{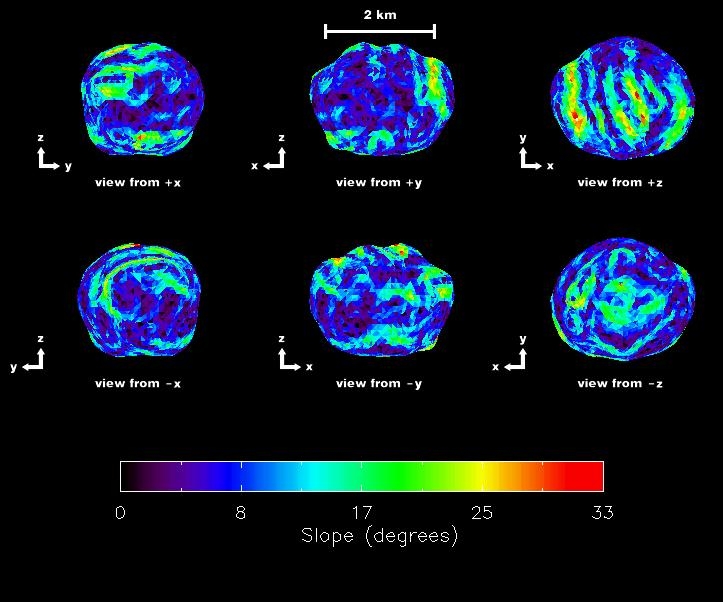}
\caption{This figure shows the gravitational slopes computed at the
center of each facet of the shape model. The gravitational slope is
the angle that the local gravitational acceleration vector makes with
the inward pointing surface-normal vector. We assumed a uniform
density of 2000~kg~m$^{-3}$ and used our measured spin period value of
8.96 hours.  Sides of the ridges near the north pole have slopes
exceeding 30$^\circ$, the approximate angle of repose of sand.
}
\label{fig:slope_map}
\end{figure*}

Fig.~\ref{fig:slope_map} shows the gravitational slope variation over
the asteroid's surface. The gravitational slope is the angle that the
local gravitational acceleration vector makes with the inward pointing
surface-normal vector. The average slope is 9.5$^\circ$.  Less than
1\% of the surface has slopes greater than 30$^\circ$, the approximate
angle of repose of sand, indicating a relaxed surface.  Slopes on the
sides of the ridges near the north pole reach up to $\sim$33$^\circ$,
and the sides facing the north pole are steeper than the opposite
sides.  As a result of the high slopes, along-the-surface
accelerations are a substantial fraction of the total acceleration,
reaching values as high as 0.34~mm~s$^{-2}$.  One might expect mass
wasting to result from this gravitational environment: the sides of
the ridges may have competent rocks exposed at the surface, and the
valley between the two ridges may be overlain by a pond of accumulated
regolith.  A similar mass wasting process is hypothesized to have
occurred on Eros~\citep{zuber00}.
The accumulation of fine-grained regolith could lower the
wavelength-scale surface roughness locally and perhaps explain the
lower values of $\mu_c$ observed at higher SRP latitudes
(Fig.~\ref{fig:cw}). 
A similar trend in $\mu_c$ was observed in the Goldstone CW spectra.

\section{Discussion}

\begin{figure}
\begin{center}

\includegraphics[scale=.33, keepaspectratio=true,angle=-90]{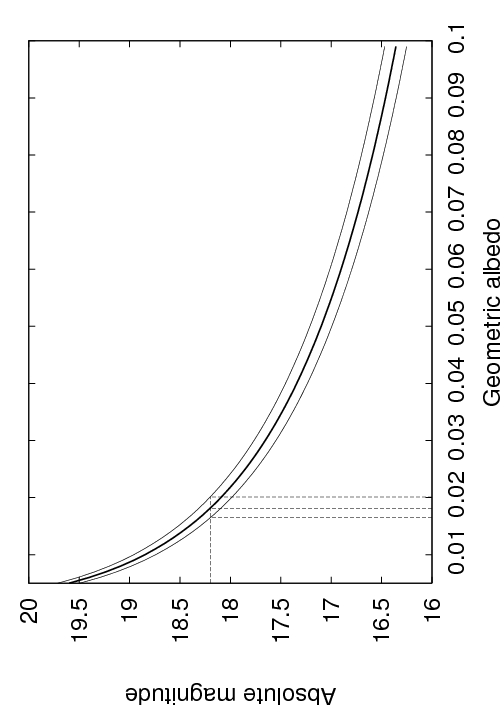}

\end{center}

\caption{Absolute magnitude vs. geometric albedo for an effective
diameter of 2.26~km (dark line) and a diameter uncertainty of 5\%
(light lines). The horizontal dashed line corresponds to the absolute
magnitude ($H$=18.2) reported by \citet{whiteley01}, and the vertical
dashed lines indicate the corresponding geometric albedos.
These albedo values are unusually low, suggesting an extraordinarily
dark object or an object having an unusual phase function.}

\label{fig:hvspv}
\end{figure}

~\citet{whiteley01} and \citet{will12} estimated 2000~ET70's absolute
magnitude ($H$) to be near 18.2. The geometric albedo ($p_V$) of an
asteroid is related to its effective diameter ($D$) and its $H$ value
by~\citep[][and references therein]{pravec07}:
\begin{equation}
p_V =\bigg[\frac{1329 {\rm \ km} \times 10^{-0.2H}}{D}\bigg]^2.
\end{equation}
Using~\citet{whiteley01}'s $H$=18.2, the above equation yields
$p_V$=0.018$\pm$0.002 for an asteroid with an effective diameter of
2.26 km $\pm$ 5\% (Fig.~\ref{fig:hvspv}). This uncertainty on $p_V$ is
due to the diameter uncertainty only. 
A geometric albedo near 2\% is extremely low compared to the albedos
of other NEAs~\citep{thomas11,stuart04}.  More common values of $p_V$
would require lower values of the absolute magnitude (e.g., $p_V$=0.04
requires $H$=17.5). The range of possible $H$ and $p_V$ values that
are consistent with the radar size estimates is shown in
Fig.~\ref{fig:hvspv}.  We conclude that 2000~ET70 has either an
extremely low albedo or unusual phase function.

~\citet{alvarez12} observed the asteroid between February 19 to 24,
when the view was close to the equator, and reported a lightcurve
amplitude of 0.60~$\pm$~0.07 mag.  If the asteroid 
was approximated by a triaxial ellipsoid with uniform albedo, this
amplitude would suggest an elongation (ratio of equatorial axes)
approximately between 1.28 and 1.36.  Our shape model 
indicates that
this ratio is $\sim$1.18, 
suggesting that either the ellipsoid approximation is poor, the
lightcurve amplitude is on the lower end of the range above, shadowing
due to the terrain is playing an important role, there are albedo
variations over the surface of the asteroid,
or a combination of these factors.

~\citet{alvarez12} also report a lightcurve period of
8.947~$\pm$~0.001 hours.  Their reported period is a function of the
intrinsic spin state of the asteroid and of the relative motion
between the asteroid, the observer, and the Sun.  Therefore, it is
close to but not exactly equivalent to the {\em synodic period}, which
combines the (fixed) intrinsic rotation and the (variable) apparent
rotation due to sky motion, but is independent of the position of the
Sun.  If we assume that the reported lightcurve period is equivalent
to the synodic period, we can compute the corresponding sidereal
periods.  This transformation depends on the spin axis orientation.
In the absence of information about the spin axis orientation, a
synodic period of 8.947 maps into sidereal periods between 8.902 h and
8.992 h, i.e., a range that is about 100 times larger than the
precision reported for the lightcurve period.  This range includes the
sidereal period that we derived from the shape modeling process
(8.960~$\pm$~0.01 h).  If we use our value and our best-fit spin axis
orientation we can evaluate corresponding synodic periods at various
epochs.  On Feb 19.0, the nominal synodic period was 8.937 hours,
whereas on Feb 25.0, the nominal synodic period was 8.943 hours.
These synodic periods are close to the reported lightcurve period, but
cannot be directly compared to it as they measure slightly different
phenomena.

Arecibo and Goldstone radar observations of 2000~ET70 allowed us
to provide a detailed characterization of a potentially hazardous
asteroid, including its size, shape, spin state, scattering
properties, and gravitational environment. These techniques are
applicable to a substantial fraction of known NEAs that make close
approaches to Earth within $\sim$0.1 AU.
Radar-based physical properties for this and other asteroids are
available at \url{http://radarastronomy.org}.

\section*{Acknowledgments}
We thank the staff at Arecibo and Goldstone for assistance with the
observations.  The Arecibo Observatory is operated by SRI
International under cooperative agreement AST-1100968 with the
National Science Foundation (NSF), and in alliance with Ana
G. M\'endez-Universidad Metropolitana, and the Universities Space
Research Association.  Some of this work was performed at the Jet
Propulsion Laboratory, California Institute of Technology, under a
contract with the National Aeronautics and Space Administration
(NASA).  This material is based in part upon work supported by NASA
under the Science Mission Directorate Research and Analysis Programs.
The Arecibo planetary radar is supported in part by NASA Near-Earth
Object Observations Program NNX12-AF24G.  SPN and JLM were partially
supported by NSF Astronomy and Astrophysics Program AST-1211581.

\bibliographystyle{plainnat}

\bibliography{et70}

\begin{thebibliography}{42}
\providecommand{\natexlab}[1]{#1}
\providecommand{\url}[1]{\texttt{#1}}
\expandafter\ifx\csname urlstyle\endcsname\relax
  \providecommand{\doi}[1]{doi: #1}\else
  \providecommand{\doi}{doi: \begingroup \urlstyle{rm}\Url}\fi

\bibitem[{Alvarez} et~al.(2012){Alvarez}, {Oey}, {Han}, {Heffner}, {Kidd},
  {Magnetta}, and {Rastede}]{alvarez12}
E.~M. {Alvarez}, J.~{Oey}, X.~L. {Han}, O.~R. {Heffner}, A.~W. {Kidd}, B.~J.
  {Magnetta}, and F.~W. {Rastede}.
\newblock {Period Determination for NEA (162421) 2000 ET70}.
\newblock \emph{Minor Planet Bulletin}, 39:\penalty0 170, July 2012.

\bibitem[{Benner} et~al.(2002){Benner}, {Ostro}, {Nolan}, {Margot}, {Giorgini},
  {Hudson}, {Jurgens}, {Slade}, {Howell}, {Campbell}, and {Yeomans}]{benn02}
L.~A.~M. {Benner}, S.~J. {Ostro}, M.~C. {Nolan}, J.~L. {Margot}, J.~D.
  {Giorgini}, R.~S. {Hudson}, R.~F. {Jurgens}, M.~A. {Slade}, E.~S. {Howell},
  D.~B. {Campbell}, and D.~K. {Yeomans}.
\newblock {Radar observations of asteroid 1999 JM8}.
\newblock \emph{Meteoritics and Planetary Science}, 37:\penalty0 779--792, June
  2002.

\bibitem[{Benner} et~al.(2006){Benner}, {Nolan}, {Ostro}, {Giorgini}, {Pray},
  {Harris}, {Magri}, and {Margot}]{benn06}
L.~A.~M. {Benner}, M.~C. {Nolan}, S.~J. {Ostro}, J.~D. {Giorgini}, D.~P.
  {Pray}, A.~W. {Harris}, C.~{Magri}, and {J.~L.} {Margot}.
\newblock {Near-Earth Asteroid 2005 CR37: Radar images and photometry of a
  candidate contact binary}.
\newblock \emph{Icarus}, 182:\penalty0 474--481, June 2006.

\bibitem[{Benner} et~al.(2008){Benner}, {Ostro}, {Magri}, {Nolan}, {Howell},
  {Giorgini}, {Jurgens}, {Margot}, {Taylor}, {Busch}, and {Shepard}]{benner08}
L.~A.~M. {Benner}, S.~J. {Ostro}, C.~{Magri}, M.~C. {Nolan}, E.~S. {Howell},
  J.~D. {Giorgini}, R.~F. {Jurgens}, J.~L. {Margot}, P.~A. {Taylor}, M.~W.
  {Busch}, and M.~K. {Shepard}.
\newblock {Near-Earth asteroid surface roughness depends on compositional
  class}.
\newblock \emph{Icarus}, 198:\penalty0 294--304, December 2008.
\newblock \doi{10.1016/j.icarus.2008.06.010}.

\bibitem[{Brozovic} et~al.(2010){Brozovic}, {Benner}, {Magri}, {Ostro},
  {Scheeres}, {Giorgini}, {Nolan}, {Margot}, {Jurgens}, and {Rose}]{broz10}
M.~{Brozovic}, L.~A.~M. {Benner}, C.~{Magri}, S.~J. {Ostro}, D.~J. {Scheeres},
  J.~D. {Giorgini}, M.~C. {Nolan}, {J.~L.} {Margot}, R.~F. {Jurgens}, and
  R.~{Rose}.
\newblock {Radar observations and a physical model of contact binary Asteroid
  4486 Mithra}.
\newblock \emph{Icarus}, 208:\penalty0 207--220, July 2010.

\bibitem[{Brozovi{\'c}} et~al.(2011){Brozovi{\'c}}, {Benner}, {Taylor},
  {Nolan}, {Howell}, {Magri}, {Scheeres}, {Giorgini}, {Pollock}, {Pravec},
  {Gal{\'a}d}, {Fang}, {Margot}, {Busch}, {Shepard}, {Reichart}, {Ivarsen},
  {Haislip}, {Lacluyze}, {Jao}, {Slade}, {Lawrence}, and {Hicks}]{broz11}
M.~{Brozovi{\'c}}, L.~A.~M. {Benner}, P.~A. {Taylor}, M.~C. {Nolan}, E.~S.
  {Howell}, C.~{Magri}, D.~J. {Scheeres}, J.~D. {Giorgini}, J.~T. {Pollock},
  P.~{Pravec}, A.~{Gal{\'a}d}, J.~{Fang}, J.~L. {Margot}, M.~W. {Busch}, M.~K.
  {Shepard}, D.~E. {Reichart}, K.~M. {Ivarsen}, J.~B. {Haislip}, A.~P.
  {Lacluyze}, J.~{Jao}, M.~A. {Slade}, K.~J. {Lawrence}, and M.~D. {Hicks}.
\newblock {Radar and optical observations and physical modeling of triple
  near-Earth Asteroid (136617) 1994 CC}.
\newblock \emph{Icarus}, 216:\penalty0 241--256, November 2011.

\bibitem[{Busch} et~al.(2006){Busch}, {Ostro}, {Benner}, {Giorgini}, {Jurgens},
  {Rose}, {Magri}, {Pravec}, {Scheeres}, and {Broschart}]{busch06}
M.~W. {Busch}, S.~J. {Ostro}, L.~A.~M. {Benner}, J.~D. {Giorgini}, R.~F.
  {Jurgens}, R.~{Rose}, C.~{Magri}, P.~{Pravec}, D.~J. {Scheeres}, and S.~B.
  {Broschart}.
\newblock {Radar and optical observations and physical modeling of near-Earth
  Asteroid 10115 (1992 SK)}.
\newblock \emph{Icarus}, 181:\penalty0 145--155, March 2006.
\newblock \doi{10.1016/j.icarus.2005.10.024}.

\bibitem[{Busch} et~al.(2008){Busch}, {Benner}, {Ostro}, {Giorgini}, {Jurgens},
  {Rose}, {Scheeres}, {Magri}, {Margot}, {Nolan}, and {Hine}]{busch08}
M.~W. {Busch}, L.~A.~M. {Benner}, S.~J. {Ostro}, J.~D. {Giorgini}, R.~F.
  {Jurgens}, R.~{Rose}, D.~J. {Scheeres}, C.~{Magri}, J.-L. {Margot}, M.~C.
  {Nolan}, and A.~A. {Hine}.
\newblock {Physical properties of near-Earth Asteroid (33342) 1998 WT24}.
\newblock \emph{Icarus}, 195:\penalty0 614--621, June 2008.
\newblock \doi{10.1016/j.icarus.2008.01.020}.

\bibitem[{Chesley} et~al.(2003){Chesley}, {Ostro}, {Vokrouhlick{\' y}}, {{\v
  C}apek}, {Giorgini}, {Nolan}, {Margot}, {Hine}, {Benner}, and
  {Chamberlin}]{ches03}
S.~R. {Chesley}, S.~J. {Ostro}, D.~{Vokrouhlick{\' y}}, D.~{{\v C}apek}, J.~D.
  {Giorgini}, M.~C. {Nolan}, J.~L. {Margot}, A.~A. {Hine}, L.~A.~M. {Benner},
  and A.~B. {Chamberlin}.
\newblock {Direct Detection of the Yarkovsky Effect by Radar Ranging to
  Asteroid 6489 Golevka}.
\newblock \emph{Science}, 302:\penalty0 1739--1742, December 2003.

\bibitem[{DeMeo} et~al.(2009){DeMeo}, {Binzel}, {Slivan}, and {Bus}]{demeo09}
F.~E. {DeMeo}, R.~P. {Binzel}, S.~M. {Slivan}, and S.~J. {Bus}.
\newblock {An extension of the Bus asteroid taxonomy into the near-infrared}.
\newblock \emph{Icarus}, 202:\penalty0 160--180, July 2009.
\newblock \doi{10.1016/j.icarus.2009.02.005}.

\bibitem[{Fang} and {Margot}(2012)]{fang12spinorbit}
J.~{Fang} and J.~L. {Margot}.
\newblock {Near-Earth Binaries and Triples: Origin and Evolution of
  Spin-Orbital Properties}.
\newblock \emph{\aj}, 143:\penalty0 24, January 2012.
\newblock \doi{10.1088/0004-6256/143/1/24}.

\bibitem[{Fang} et~al.(2011){Fang}, {Margot}, {Brozovic}, {Nolan}, {Benner},
  and {Taylor}]{fang11triples}
J.~{Fang}, J.~L. {Margot}, M.~{Brozovic}, M.~C. {Nolan}, L.~A.~M. {Benner}, and
  P.~A. {Taylor}.
\newblock {Orbits of Near-Earth Asteroid Triples 2001 SN263 and 1994 CC:
  Properties, Origin, and Evolution}.
\newblock \emph{\aj}, 141:\penalty0 154--+, May 2011.

\bibitem[{Hudson} and {Ostro}(1999)]{hudson99}
R.~S. {Hudson} and S.~J. {Ostro}.
\newblock {Physical Model of Asteroid 1620 Geographos from Radar and Optical
  Data}.
\newblock \emph{Icarus}, 140:\penalty0 369--378, August 1999.
\newblock \doi{10.1006/icar.1999.6142}.

\bibitem[{Hudson} et~al.(2000){Hudson}, {Ostro}, {Jurgens}, {Rosema},
  {Giorgini}, {Winkler}, {Rose}, {Choate}, {Cormier}, {Franck}, {Frye},
  {Howard}, {Kelley}, {Littlefair}, {Slade}, {Benner}, {Thomas}, {Mitchell},
  {Chodas}, {Yeomans}, {Scheeres}, {Palmer}, {Zaitsev}, {Koyama}, {Nakamura},
  {Harris}, and {Meshkov}]{hudson00}
R.~S. {Hudson}, S.~J. {Ostro}, R.~F. {Jurgens}, K.~D. {Rosema}, J.~D.
  {Giorgini}, R.~{Winkler}, R.~{Rose}, D.~{Choate}, R.~A. {Cormier}, C.~R.
  {Franck}, R.~{Frye}, D.~{Howard}, D.~{Kelley}, R.~{Littlefair}, M.~A.
  {Slade}, L.~A.~M. {Benner}, M.~L. {Thomas}, D.~L. {Mitchell}, P.~W. {Chodas},
  D.~K. {Yeomans}, D.~J. {Scheeres}, P.~{Palmer}, A.~{Zaitsev}, Y.~{Koyama},
  A.~{Nakamura}, A.~W. {Harris}, and M.~N. {Meshkov}.
\newblock {Radar Observations and Physical Model of Asteroid 6489 Golevka}.
\newblock \emph{Icarus}, 148:\penalty0 37--51, November 2000.
\newblock \doi{10.1006/icar.2000.6483}.

\bibitem[{Hudson} and {Ostro}(1994)]{huds94}
R.S. {Hudson} and S.J. {Ostro}.
\newblock Shape of asteroid 4769 {Castalia (1989 PB)} from inversion of radar
  images.
\newblock \emph{Science}, 263:\penalty0 940--943, February 1994.

\bibitem[{Hudson}(1993)]{hudson93}
S.~{Hudson}.
\newblock {Three-dimensional reconstruction of asteroids from radar
  observations}.
\newblock \emph{Remote Sensing Reviews}, 8:\penalty0 195--203, 1993.

\bibitem[{La Spina} et~al.(2004){La Spina}, {Paolicchi}, {Kryszczy{\'n}ska},
  and {Pravec}]{spina04}
A.~{La Spina}, P.~{Paolicchi}, A.~{Kryszczy{\'n}ska}, and P.~{Pravec}.
\newblock {Retrograde spins of near-Earth asteroids from the Yarkovsky effect}.
\newblock \emph{\nat}, 428:\penalty0 400--401, March 2004.
\newblock \doi{10.1038/nature02411}.

\bibitem[{Lowry} et~al.(2007){Lowry}, {Fitzsimmons}, {Pravec},
  {Vokrouhlick{\'y}}, {Boehnhardt}, {Taylor}, {Margot}, {Gal{\'a}d}, {Irwin},
  {Irwin}, and {Kusnir{\'a}k}]{lowr07}
S.~C. {Lowry}, A.~{Fitzsimmons}, P.~{Pravec}, D.~{Vokrouhlick{\'y}},
  H.~{Boehnhardt}, P.~A. {Taylor}, J.~L. {Margot}, A.~{Gal{\'a}d}, M.~{Irwin},
  J.~{Irwin}, and P.~{Kusnir{\'a}k}.
\newblock {Direct Detection of the Asteroidal YORP Effect}.
\newblock \emph{Science}, 316:\penalty0 272--, April 2007.

\bibitem[{Magri} et~al.(2007){Magri}, {Ostro}, {Scheeres}, {Nolan}, {Giorgini},
  {Benner}, and {Margot}]{magri07}
C.~{Magri}, S.~J. {Ostro}, D.~J. {Scheeres}, M.~C. {Nolan}, J.~D. {Giorgini},
  L.~A.~M. {Benner}, and J.~L. {Margot}.
\newblock {Radar observations and a physical model of Asteroid 1580 Betulia}.
\newblock \emph{Icarus}, 186:\penalty0 152--177, January 2007.
\newblock \doi{10.1016/j.icarus.2006.08.004}.

\bibitem[Margot(2001)]{marg01chirp}
J.~L. Margot.
\newblock {Planetary Radar Astronomy with Linear FM (chirp) Waveforms}.
\newblock Arecibo technical and operations memo series 2001-09, Arecibo
  Observatory, 2001.

\bibitem[{Margot} et~al.(2002){Margot}, {Nolan}, {Benner}, {Ostro}, {Jurgens},
  {Giorgini}, {Slade}, and {Campbell}]{margot02}
J.~L. {Margot}, M.~C. {Nolan}, L.~A.~M. {Benner}, S.~J. {Ostro}, R.~F.
  {Jurgens}, J.~D. {Giorgini}, M.~A. {Slade}, and D.~B. {Campbell}.
\newblock {Binary Asteroids in the Near-Earth Object Population}.
\newblock \emph{Science}, 296:\penalty0 1445--1448, May 2002.
\newblock \doi{10.1126/science.1072094}.

\bibitem[{Mitchell} et~al.(1996){Mitchell}, {Ostro}, {Hudson}, {Rosema},
  {Campbell}, {Velez}, {Chandler}, {Shapiro}, {Giorgini}, and
  {Yeomans}]{mitchell96}
D.~L. {Mitchell}, S.~J. {Ostro}, R.~S. {Hudson}, K.~D. {Rosema}, D.~B.
  {Campbell}, R.~{Velez}, J.~F. {Chandler}, I.~I. {Shapiro}, J.~D. {Giorgini},
  and D.~K. {Yeomans}.
\newblock {Radar Observations of Asteroids 1 Ceres, 2 Pallas, and 4 Vesta}.
\newblock \emph{Icarus}, 124:\penalty0 113--133, November 1996.
\newblock \doi{10.1006/icar.1996.0193}.

\bibitem[{Nolan} et~al.(2008){Nolan}, {Howell}, {Becker}, {Magri}, {Giorgini},
  and {Margot}]{nola08dps}
M.~C. {Nolan}, E.~S. {Howell}, T.~M. {Becker}, C.~{Magri}, J.~D. {Giorgini},
  and J.~L. {Margot}.
\newblock {Arecibo Radar Observations of 2001 SN263: A Near-Earth Triple
  Asteroid System}.
\newblock In \emph{Bulletin of the American Astronomical Society}, volume~40,
  2008.

\bibitem[Nugent et~al.(2012)Nugent, Margot, Chesley, and
  Vokrouhlick{\'y}]{nuge12yark}
C.~R. Nugent, J.~L. Margot, S.~R. Chesley, and D.~Vokrouhlick{\'y}.
\newblock Detection of semi-major axis drifts in 54 near-earth asteroids: New
  measurements of the yarkovsky effect.
\newblock \emph{Astronomical Journal}, 144:\penalty0 60, 2012.
\newblock Arxiv eprint 1204.5990.

\bibitem[{Ostro}(1993)]{ostro93}
S.~J. {Ostro}.
\newblock {Planetary radar astronomy}.
\newblock \emph{Reviews of Modern Physics}, 65:\penalty0 1235--1279, October
  1993.
\newblock \doi{10.1103/RevModPhys.65.1235}.

\bibitem[{Ostro} et~al.(1983){Ostro}, {Campbell}, and {Shapiro}]{ostro83}
S.~J. {Ostro}, D.~B. {Campbell}, and I.~I. {Shapiro}.
\newblock {Radar observations of asteroid 1685 Toro}.
\newblock \emph{\aj}, 88:\penalty0 565--576, April 1983.
\newblock \doi{10.1086/113345}.

\bibitem[{Ostro} et~al.(1995){Ostro}, {Hudson}, {Jurgens}, {Rosema},
  {Campbell}, {Yeomans}, {Chandler}, {Giorgini}, {Winkler}, {Rose}, {Howard},
  {Slade}, {Perillat}, and {Shapiro}]{ostro95}
S.~J. {Ostro}, R.~S. {Hudson}, R.~F. {Jurgens}, K.~D. {Rosema}, D.~B.
  {Campbell}, D.~K. {Yeomans}, J.~F. {Chandler}, J.~D. {Giorgini},
  R.~{Winkler}, R.~{Rose}, S.~D. {Howard}, M.~A. {Slade}, P.~{Perillat}, and
  I.~I. {Shapiro}.
\newblock {Radar Images of Asteroid 4179 Toutatis}.
\newblock \emph{Science}, 270:\penalty0 80--83, October 1995.
\newblock \doi{10.1126/science.270.5233.80}.

\bibitem[{Ostro} et~al.(2001){Ostro}, {Hudson}, {Benner}, {Nolan}, {Giorgini},
  {Scheeres}, {Jurgens}, and {Rose}]{ostro01}
S.~J. {Ostro}, R.~S. {Hudson}, L.~A.~M. {Benner}, M.~C. {Nolan}, J.~D.
  {Giorgini}, D.~J. {Scheeres}, R.~F. {Jurgens}, and R.~{Rose}.
\newblock {Radar observations of asteroid 1998 ML14}.
\newblock \emph{Meteoritics and Planetary Science}, 36:\penalty0 1225--1236,
  September 2001.
\newblock \doi{10.1111/j.1945-5100.2001.tb01956.x}.

\bibitem[{Ostro} et~al.(2006){Ostro}, {Margot}, {Benner}, {Giorgini},
  {Scheeres}, {Fahnestock}, {Broschart}, {Bellerose}, {Nolan}, {Magri},
  {Pravec}, {Scheirich}, {Rose}, {Jurgens}, {De Jong}, and {Suzuki}]{ostro06}
S.~J. {Ostro}, J.~L. {Margot}, L.~A.~M. {Benner}, J.~D. {Giorgini}, D.~J.
  {Scheeres}, E.~G. {Fahnestock}, S.~B. {Broschart}, J.~{Bellerose}, M.~C.
  {Nolan}, C.~{Magri}, P.~{Pravec}, P.~{Scheirich}, R.~{Rose}, R.~F. {Jurgens},
  E.~M. {De Jong}, and S.~{Suzuki}.
\newblock {Radar Imaging of Binary Near-Earth Asteroid (66391) 1999 KW4}.
\newblock \emph{Science}, 314:\penalty0 1276--1280, November 2006.
\newblock \doi{10.1126/science.1133622}.

\bibitem[Peebles(2007)]{peebles07}
P.Z. Peebles.
\newblock \emph{Radar Principles}.
\newblock Wiley India Pvt. Limited, 2007.
\newblock ISBN 9788126515271.
\newblock URL \url{http://books.google.com/books?id=rnX21aAMKCIC}.

\bibitem[{Pravec} and {Harris}(2007)]{pravec07}
P.~{Pravec} and A.~W. {Harris}.
\newblock {Binary asteroid population. 1. Angular momentum content}.
\newblock \emph{Icarus}, 190:\penalty0 250--259, September 2007.
\newblock \doi{10.1016/j.icarus.2007.02.023}.

\bibitem[{Scheeres} et~al.(2006){Scheeres}, {Fahnestock}, {Ostro}, {Margot},
  {Benner}, {Broschart}, {Bellerose}, {Giorgini}, {Nolan}, {Magri}, {Pravec},
  {Scheirich}, {Rose}, {Jurgens}, {De Jong}, and {Suzuki}]{scheeres06}
D.~J. {Scheeres}, E.~G. {Fahnestock}, S.~J. {Ostro}, J.~L. {Margot}, L.~A.~M.
  {Benner}, S.~B. {Broschart}, J.~{Bellerose}, J.~D. {Giorgini}, M.~C. {Nolan},
  C.~{Magri}, P.~{Pravec}, P.~{Scheirich}, R.~{Rose}, R.~F. {Jurgens}, E.~M.
  {De Jong}, and S.~{Suzuki}.
\newblock {Dynamical Configuration of Binary Near-Earth Asteroid (66391) 1999
  KW4}.
\newblock \emph{Science}, 314:\penalty0 1280--1283, November 2006.
\newblock \doi{10.1126/science.1133599}.

\bibitem[{Shepard} et~al.(2006){Shepard}, {Margot}, {Magri}, {Nolan},
  {Schlieder}, {Estes}, {Bus}, {Volquardsen}, {Rivkin}, {Benner}, {Giorgini},
  {Ostro}, and {Busch}]{shepard06}
M.~K. {Shepard}, J.~L. {Margot}, C.~{Magri}, M.~C. {Nolan}, J.~{Schlieder},
  B.~{Estes}, S.~J. {Bus}, E.~L. {Volquardsen}, A.~S. {Rivkin}, L.~A.~M.
  {Benner}, J.~D. {Giorgini}, S.~J. {Ostro}, and M.~W. {Busch}.
\newblock {Radar and infrared observations of binary near-Earth Asteroid 2002
  CE26}.
\newblock \emph{Icarus}, 184:\penalty0 198--210, September 2006.
\newblock \doi{10.1016/j.icarus.2006.04.019}.

\bibitem[{Stuart} and {Binzel}(2004)]{stuart04}
J.~S. {Stuart} and R.~P. {Binzel}.
\newblock {Bias-corrected population, size distribution, and impact hazard for
  the near-Earth objects}.
\newblock \emph{Icarus}, 170:\penalty0 295--311, August 2004.
\newblock \doi{10.1016/j.icarus.2004.03.018}.

\bibitem[{Taylor} and {Margot}(2011)]{tayl11}
P.~A. {Taylor} and J.~L. {Margot}.
\newblock {Binary asteroid systems: Tidal end states and estimates of material
  properties}.
\newblock \emph{Icarus}, 212:\penalty0 661--676, April 2011.

\bibitem[{Taylor} et~al.(2007){Taylor}, {Margot}, {Vokrouhlick{\'y}},
  {Scheeres}, {Pravec}, {Lowry}, {Fitzsimmons}, {Nolan}, {Ostro}, {Benner},
  {Giorgini}, and {Magri}]{tayl07}
P.~A. {Taylor}, J.~L. {Margot}, D.~{Vokrouhlick{\'y}}, D.~J. {Scheeres},
  P.~{Pravec}, S.~C. {Lowry}, A.~{Fitzsimmons}, M.~C. {Nolan}, S.~J. {Ostro},
  L.~A.~M. {Benner}, J.~D. {Giorgini}, and C.~{Magri}.
\newblock {Spin Rate of Asteroid (54509) 2000 PH5 Increasing Due to the YORP
  Effect}.
\newblock \emph{Science}, 316:\penalty0 274--, April 2007.
\newblock \doi{10.1126/science.1139038}.

\bibitem[{Tholen}(1984)]{tholen84}
D.~J. {Tholen}.
\newblock \emph{{Asteroid taxonomy from cluster analysis of Photometry}}.
\newblock PhD thesis, Arizona Univ., Tucson., 1984.

\bibitem[{Thomas} et~al.(2011){Thomas}, {Trilling}, {Emery}, {Mueller}, {Hora},
  {Benner}, {Bhattacharya}, {Bottke}, {Chesley}, {Delb{\'o}}, {Fazio},
  {Harris}, {Mainzer}, {Mommert}, {Morbidelli}, {Penprase}, {Smith}, {Spahr},
  and {Stansberry}]{thomas11}
C.~A. {Thomas}, D.~E. {Trilling}, J.~P. {Emery}, M.~{Mueller}, J.~L. {Hora},
  L.~A.~M. {Benner}, B.~{Bhattacharya}, W.~F. {Bottke}, S.~{Chesley},
  M.~{Delb{\'o}}, G.~{Fazio}, A.~W. {Harris}, A.~{Mainzer}, M.~{Mommert},
  A.~{Morbidelli}, B.~{Penprase}, H.~A. {Smith}, T.~B. {Spahr}, and J.~A.
  {Stansberry}.
\newblock {ExploreNEOs. V. Average Albedo by Taxonomic Complex in the
  Near-Earth Asteroid Population}.
\newblock \emph{\aj}, 142:\penalty0 85, September 2011.
\newblock \doi{10.1088/0004-6256/142/3/85}.

\bibitem[{Werner} and {Scheeres}(1997)]{werner97}
R.~A. {Werner} and D.~J. {Scheeres}.
\newblock {Exterior Gravitation of a Polyhedron Derived and Compared with
  Harmonic and Mascon Gravitation Representations of Asteroid 4769 Castalia}.
\newblock \emph{Celestial Mechanics and Dynamical Astronomy}, 65:\penalty0
  313--344, 1997.

\bibitem[{Whiteley}(2001)]{whiteley01}
R.~J. {Whiteley}, Jr.
\newblock \emph{{A compositional and dynamical survey of the near-Earth
  asteroids}}.
\newblock PhD thesis, University of Hawai'i at Manoa, 2001.

\bibitem[{Williams}(2012)]{will12}
G.~V. {Williams}.
\newblock \emph{{Minor Planet Astrophotometry}}.
\newblock PhD thesis, Smithsonian Astrophysical Observatory, 2012.
\newblock {gwilliams@cfa.harvard.edu}.

\bibitem[{Zuber} et~al.(2000){Zuber}, {Smith}, {Cheng}, {Garvin}, {Aharonson},
  {Cole}, {Dunn}, {Guo}, {Lemoine}, {Neumann}, {Rowlands}, and
  {Torrence}]{zuber00}
M.~T. {Zuber}, D.~E. {Smith}, A.~F. {Cheng}, J.~B. {Garvin}, O.~{Aharonson},
  T.~D. {Cole}, P.~J. {Dunn}, Y.~{Guo}, F.~G. {Lemoine}, G.~A. {Neumann}, D.~D.
  {Rowlands}, and M.~H. {Torrence}.
\newblock {The Shape of 433 Eros from the NEAR-Shoemaker Laser Rangefinder}.
\newblock \emph{Science}, 289:\penalty0 2097--2101, September 2000.
\newblock \doi{10.1126/science.289.5487.2097}.

\end{thebibliography}

\end{document}